\documentclass[11pt]{article}
\usepackage{amsfonts}
\usepackage{amsmath}
 \usepackage{amssymb}
 \newfont{\bbbold}{msbm10}

 \def\bbR{\mbox{\bbbold R}}

 \def\cL{{\cal L}}

 \newfont{\goth}{eufm10 scaled \magstep1}

 \def\a{\alpha}
 \def\b{\beta}
 \def\c{\gamma}\def\C{\Gamma}
 \def\d{\delta}
 \def\e{\epsilon}
 \def\f{\phi}\def\vf{\varphi}
 \def\h{\eta}

 \def\l{\lambda}
 \def\m{\mu}

 \def\t{\tau}
 \def\th{\theta}

 \def\be{\begin{equation}}\def\ee{\end{equation}}
 \def\bea{\begin{eqnarray}}\def\eea{\end{eqnarray}}
 \def\ba{\begin{array}}\def\ea{\end{array}}
 
 \def\o{\omega}\def\O{\Omega}
 \def\del{\partial}

 \def\str{\rm str}

 \def\nab{\nabla}

 \def\del{\partial}

 \def\3dt{\dot{3}}
 
\def\half{{1\over2}}
\def\To{\stackrel{0}{T}}
\def\Ti{\stackrel{1}{T}}
\def\Fo{\stackrel{0}{F}}
\def\Go{\stackrel{0}{G}}\def\Gi{\stackrel{1}{G}}
\def\tao{\stackrel{0}{\t}}\def\tai{\stackrel{1}{\t}}

 \let\la=\label

 \let\bm=\bibitem{}

 \def\nn{\nonumber}
 \def\bd{\begin{document}}
 \def\ed{\end{document}}
 \def\bea{\begin{eqnarray}}
 \def\ba{\begin{array}}\def\ea{\end{array}}
 \def\eea{\end{eqnarray}}
 \def\ft#1#2{{\textstyle{{\scriptstyle #1}\over {\scriptstyle #2}}}}
 \def\fft#1#2{{#1 \over #2}}
 \newcommand{\eq}[1]{(\ref{#1})}
 \def\eqs#1#2{(\ref{#1}-\ref{#2})}
 \def\det{{\rm det\,}}
 \def\tr{{\rm tr}}\def\Tr{{\rm Tr}}
  \def\str{{\rm str}} \def\diag{{\rm diag}}
 \def\sdet{{\rm sdet}}
 
\begin{document}

 \thispagestyle{empty}

 \hfill{KCL-TH-03-02}

  \hfill{\today}

 \vspace{20pt}

 \begin{center}
 {\Large{\bf On higher-order corrections in M theory}}
 \vspace{30pt}

 {P.S. Howe and D. Tsimpis} \vskip 1cm {Department of Mathematics}
 \vskip 1cm {King's College, London} \vspace{15pt}

 \vspace{60pt}

 \end{center}

 {\bf Abstract}
A theoretical analysis of higher-order corrections to $D=11$
supergravity is given in a superspace framework. It is shown that
any deformation of $D=11$ supergravity for which the
lowest-dimensional component of the four-form $G_4$ vanishes is
trivial. This implies that the equations of motion of $D=11$
supergravity are specified by an element of a certain spinorial
cohomology group and generalises previous results obtained using
spinorial or pure spinor cohomology  to the fully non-linear
theory. The first deformation of the theory is given by an element
of a different spinorial cohomology group with coefficients which
are local tensorial functions of the massless supergravity fields.
The four-form Bianchi  Identities are solved, to first order and
at dimension $-\frac{1}{2}$, in the case that the
lowest-dimensional component of $G_4$ is non-zero. Moreover, it is
shown how one can calculate the first-order correction to the
dimension-zero torsion and thus to the supergravity equations of
motion given an explicit expression for this object in terms of
the supergravity fields. The version of the theory with both a
four-form and a seven-form is discussed in the presence of the
five-brane anomaly-cancelling term. It is shown that the
supersymmetric completion of this term exists and it is argued
that it is the unique anomaly-cancelling invariant at this
dimension which is at least quartic in the fields.  This implies
that the first deformation of the theory is completely determined
by the anomaly term from which one can, in principle, read off the
corrections to all of the superspace field strength tensors.

 {\vfill\leftline{}\vfill \vskip  10pt

 \baselineskip=15pt \pagebreak \setcounter{page}{1}

\section{Introduction}

Although $M$-theory has been around for some time now, it is still
not clear precisely what the theory is, although there have been
many suggestions \cite{Duff:ys}. In many situations, however, it
would be sufficient to know something about the corrections to the
effective theory of massless fields. In the absence of an
underlying perturbative analogue of string theory such corrections
cannot be found systematically, even in principle, but we might
hope that supersymmetry would be sufficient to determine the
first-order corrections and perhaps something beyond that. This is
the subject matter of the present paper: we show that, by taking
the lowest-dimensional component of the closed superspace
four-form $G_4$ to vanish, we automatically arrive at the standard
supergravity equations. This result can be rephrased in terms of a
certain spinorial cohomology group and is the generalisation to
the non-linear theory of the result proved at the linearised level
by CNT \cite{cntc} and Berkovits in \cite{Berkovits:2002uc}. We
also show that the first deformation to the lowest-order theory is
given by an element of another  spinorial cohomology group,
although with restricted coefficients.  We go on to solve the
four-form Bianchi identities, to first order and at dimension
one-half, given an arbitrary non-zero $G_{\a\b\c\d}$. In
particular, we show how one can construct the first-order
correction to the dimension-zero component of the torsion tensor
in this fashion.  We also show that this torsion automatically
satisfies the constraints imposed by the first geometrical Bianchi
identity. We then comment on the four-form/seven-form formulation
of the theory in the presence of anomalies. We construct (in
principle) an invariant which includes the fivebrane
anomaly-cancelling Chern-Simons term and argue that this is the
only invariant, apart from the classical action, which has the
same or lower dimension than $R^4$ and which is at least quartic
in the fields. It is also argued that, at this order, all the
components of the superspace field strength tensors can be
determined systematically using the Bianchi identities and the
input of the anomaly term.

There have been many papers on higher-order corrections in
supergravity and string theory. Some relevant papers are
\cite{Green:1997di,Green:1997as,Green:1997me,Antoniadis:1997eg,
Kiritsis:1997em,Berkovits:1998ex,Green:1998by,Tseytlin:2000sf} and
\cite{Peeters:2000qj} the last of which contains a review and
further references. However, the construction of the complete
$R^4$ terms in type II string theory or in M-theory is (and
remains) a difficult problem.  A recent attempt was made in
\cite{deHaro:2002vk} to construct an $R^4$ action in type IIB
based on a chiral measure in on-shell IIB superspace, but it was
subsequently observed that such a measure does not exist
\cite{bh}. A completely different recent approach was taken by the
authors of \cite{Peeters:2000qj} who used partial results from
type II string theory and attempted to lift them to eleven
dimensions, but this proved to be  too difficult to carry out
completely.

There have also been several papers on the subject directly in
$D=11$ superspace. We recall that the equations of motion of
$D=11$ supergravity, discovered in \cite{Cremmer:1978km}, were
recast in superspace form in references
\cite{Cremmer:1980ru,Brink:1980az}. In the first of these, the
superspace constraints were constructed from the component theory
by the method of gauge completion, while in the second, they were
found from the supergravity multiplet by dimensional analysis. At
the time no attempt was made to extract a minimal set of
constraints, but there have subsequently been a number of papers
on this subject. In \cite{Howe:1991bx} one of the present authors
showed that the equations of motion could be understood as
integrability conditions for a certain differential operator in
membrane superspace (the space of maps from a two-dimensional
spatial membrane to the $D=11$ superspace). This approach makes
use of pure spinors, i.e. complex bosonic spinors $u^{\a}$
satisfying

\be u^{\a} (\c^a)_{\a\b}u^{\b}=u^{\a} (\c^{ab})_{\a\b}u^{\b} =0
\la{1.1} \ee

The integrability conditions lead to constraints on both
$T_{\a\b}{}^c$ and $G_{\a\b CD}$ which together imply the
equations of motion, although again this is a redundant set.
Explicitly,  the constraints found this way are

 \be
 (\c_{abcde})^{\a\b} T_{\a\b}{}^f=(\c_{abcde})^{\a\b} G_{\a\b
 CD}=0
 \ee
Somewhat later, it was shown that it is only necessary to consider
the geometrical part of the theory and to set

\be T_{\a\b}{}^c=-i(\c^c)_{\a\b} \la{1.2} \ee

to obtain the supergravity equations of motion \cite{Howe:1997rf}.
This result is most easily established in Weyl superspace in which
the structure group is taken to include a scale factor.

On the basis of this information an extensive analysis of the
geometrical Bianchi identities was carried out in \cite{cgnn} for
the case of a partially-modified $T_{\a\b}{}^c$. A complete
investigation is in progress \cite{cgnt}. A discussion of the
relevant spinorial cohomology groups with unrestricted
coefficients was given in \cite{cntc}. In \cite{cgnn}
it was observed that, if the negative dimension components of
$G_{ABCD}$ are taken to vanish, then the resulting theory cannot
generate M-theory corrections. This point was discussed in more
detail in \cite{Nishino:2001mb}.

In a recent paper, in which he attempted to generalise his pure
spinor approach to string theory to the supermembrane, Berkovits
\cite{Berkovits:2002uc} discussed the quantisation of the $D=11$
superparticle using a BRST operator of the form $Q:=u^{\a} D_{\a}$
where $u^{\a}$ is taken to satisfy only the first constraint of
\eq{1.1},  i.e. $u\c^a u=0$, and where $D_{\a}$ is the flat
superspace derivative.\footnote{We shall refer to such an object
as a B-pure spinor to distinguish it from a pure spinor in
Cartan's sense which obeys \eq{1.1} in $D=11$.} He showed that the
linearised equations of motion of $D=11$ supergravity are
equivalent  to finding an element of the ghost number three
cohomology group of $Q$ where $u$ is assigned ghost number one.
One of the results obtained in the present paper (section four) is
a proof that any deformation of the $D=11$ supergravity equations
of motion for which $G_{\a\b\c\d}=0$ is trivial which is
essentially a non-linear reformulation of the CNT/Berkovits
result. To see this one needs to know a little bit about spinorial
cohomology and  this is discussed in a geometrical framework in
the next section.  The notion of spinorial cohomology was
introduced in \cite{cnta, cntc} and was used in explicit
investigations of ten-dimensional SYM in \cite{cntb, cntd}. In
these references it was pointed out that supersymmetric
deformations can be understood perturbatively as elements in the
cohomology of an appropriately defined nilpotent spinor operator.
The latter was expressed in terms of a supersymmetic spinor
derivative followed by a projection onto the highest-weight
representation.  The relation between spinorial and pure spinor
cohomologies is discussed in Appendix D.

In section four we consider the $dG_4=0$ Bianchi identities in the
case that $G_{\a\b\c\d}\neq 0$ and show how one may obtain the
dimension zero component of the torsion tensor in terms of
derivatives of this object by this means. Further details of the
calculation are given in section six. In section five we discuss
what happens when the seven-form field strength is included. We
show how to construct an action which is fully supersymmetric and
which includes the Chern-Simons term necessary to cancel the
fivebrane anomaly. Appendices A, B and C cover some technical
details of the four-form Bianchi identity calculations.


\section{Supergravity}

We recall that $D=11$ supergravity can be described \`a la
Wess-Zumino in a superspace with eleven even and thirty-two odd
dimenions in which the structure group is taken to be the
eleven-dimensional spin group. The preferred coframes are denoted
$E^A=(E^a,E^{\a})$ where $a=0,\ldots 10$ is a  Lorentz vector
index and $\a=1,\ldots 32$ is a spinor index, the spinors being
Majorana. The dual frame basis of vector fields is denoted by
$E_A=(E_a, E_{\a})$. The preferred coframes are related to the
coordinate basis by the supervielbein, $E^A=dz^M E_M{}^A$. Note
that the splitting of the tangent space into even and odd is
respected by the structure group. We also introduce a connection
one-form $\O_A{}^B$ which takes its values in the Lie algebra of
the Lorentz group and define the torsion and curvature in the
usual way:

\bea
T^A &=&DE^A:=dE^A + E^B \O_B{}^A=\half E^C E^B T_{BC}{}^A \nn\\
R_A{}^B &=&d\O_A{}^B +\O_A{}^C\O_C{}^B=\half E^D E^C R_{CD,A}{}^B
\la{2.1}
 \eea

{}From the definitions we have the Bianchi identities $DT^A=E^B
R_B{}^A$ and $DT^A=0$. The assumption that the structure group is
the Lorentz group implies that $R_a{}^{\b}=R_{\a}{}^b=0$ while

\be R_{\a}{}^{\b}={1\over4} (\c^{ab})_{\a}{}^{\b} R_{ab} \ee

We also require that we have supersymmetry, so that the dimension
zero torsion tensor should have the form

\be T_{\a\b}{}^c=-i\left ((\c^c)_{\a\b} + (\c^{d_1 d_2})_{\a\b}
X_{d_1 d_2}{}^c + (\c^{d_1 d_2 d_3 d_4 d_5})_{\a\b} Y_{d_1 d_2 d_3
d_4 d_5}{}^c\right) \la{2.2}\ee

where the tensors $X$ and $Y$ are traceless and have vanishing
totally antisymmetric parts. Note that these constraints on
$T_{\a\b}{}^c$ can be imposed by suitable field redefinitions of
the supervielbein \cite{Howe:1997wf}.

Given this basic structure Dragon's theorem \cite{Dragon:nf}
implies that the components of the curvature tensor can be
computed from the components of the torsion tensor from the first
Bianchi identity while the second Bianchi identity is
automatically satisfied if the first one is. In index notation the
first identity reads

 \be
 I_{ABC}{}^D:=\nab_{[A} T_{BC]}{}^D + T_{[AB}{}^E T_{|E|C]}{}^D -
 R_{[AB,C]}{}^D
 \la{2.3}
 \ee

where the square brackets denote graded antisymmetrisation.

If we take the dimension-zero tensors $X$ and $Y$ to be zero, then
the Bianchi identities imply that there is a single spinor
superfield at dimension one-half after all the permissible field
redefinitions have been carried out \cite{Candiello:di}. However,
this superfield is in fact a derivative of a scalar superfield
which can be removed by a rescaling of the supervielbein. This can
be most easily seen by including scale transformations in the
structure group and modifying the connection and curvature
\cite{Howe:1997rf}. One can then show that taking the dimension
zero torsion to be $-i(\c^c)_{\a\b}$ implies that there are no
dimension one-half fields (because we have the freedom to choose
an extra conventional constraint corresponding to the scale part
of the connection), and furthermore that the scale part of the
curvature vanishes. We can therefore set the scale connection to
zero and return to a Lorentzian superspace which will now have no
dimension one-half field present.  The fields that one is left
with are those of the on-shell supergravity theory which must (and
do) obey their equations of motion by virtue of supersymmetry.

We recall that these fields are the graviton, the gravitino and a
three-form potential. Since these are all gauge fields they are
visible in the components of the torsion tensor through their
field strengths. The field strengths of the gravitino and
curvature can be identified with the leading components of the
superspace torsion tensor at dimension three-halves,
$T_{ab}{}^{\c}$, and the curvature tensor at dimension two,
$R_{ab,c}{}^d$, respectively. The dimension-one field strength of
the three-form potential, which we shall denote by $W_{abcd}$,
appears in the dimension-one components of both the torsion tensor
and the curvature:

\be T_{a\b}{}^{\c}= -{1\over36}\left((\c^{bcd})_{\b}{}^{\c}
W_{abcd} +{1\over8} (\c_{abcde})_{\b}{}^{\c} W^{abcd}\right) \ee

and

\be R_{\a\b,ab} ={i\over 6}\left((\c^{cd})_{\a\b} W_{abcd} +
{1\over24}(\c_{abcdef})_{\a\b} W^{cdef}\right) \ee

Furthermore, the independent components of the superfield $W$ are
precisely the field strengths of the component fields of the
supergravity multiplet.

At this point, we have not located the three-form potential, but
we can construct a closed superspace four-form $G_4$ whose only
non-vanishing components are $G_{abcd}=W_{abcd}$ and

\be G_{\a\b cd}=-i(\c_{cd})_{\a\b} \ee

The fact that this $G_4$ satisfies $dG_4=0$ does not require any
new information that we have not met before in the analysis of the
Bianchi identities for the torsion and curvature. We therefore
deduce the existence of a three-form potential $C$ such that
$G_4=dC$.

To summarise, then,  we have seen that the equations of motion of
$D=11$ supergravity, as well as the component field content of the
theory, follow from the constraint $T_{\a\b}{}^c=-i(\c^c)_{\a\b}$.
Furthermore, if we were to take a minimalist point of view, we
could in principle dispense with the connection formalism and just
consider $T_{\a\b}{}^c$ on its own.  However, we shall not pursue
this approach since it is more convenient to use the usual
connection formalism especially when one wishes to consider
higher-order corrections.


\section{Spinorial cohomology}

In this section we shall give a geometrical definition of
spinorial cohomology. In order to define such a concept one has to
choose an odd tangent bundle $F$. We shall also suppose that we
have the usual machinery of supergravity, i.e. Lorentzian
structure group, connection, torsion and curvature as discussed
above although it is possible to avoid much of this. In particular
we suppose that the tangent bundle $T$ is a direct sum of the odd
and even bundles. In making a choice of the complementary even
sub-bundle $B$ we are free to fix a conventional constraint on the
dimension one-half torsion. A natural choice would be

 \be
 (\c_a)^{\a\b} T_{\a\b}{}^\c=0
 \la{3.1}
 \ee

The space of forms admits a natural bigrading according to the
degrees of the forms and their Grassmann character. The space of
forms with $p$ even and $q$ odd components is denoted by
$\O^{p,q}$; $\o\in\O^{p,q}$ has the form

 \be
 \o=E^{\b_q}\ldots E^{\b_1}\, E^{a_p}\ldots E^{a_1} \o_{a_1\ldots
 a_p \b_1\ldots \b_q}
 \la{3.2}
 \ee

The exterior derivative $d$ written out in this basis naturally
involves the components of the torsion as one has to differentiate
the one-form basis elements. It maps $\O^{p,q}$ to
$\O^{p+1,q}+\O^{p,q+1} +\O^{p-1,q+2}+\O^{p+2,q-1}$. Following
\cite{Bonora:mt} we split $d$ into its various components with
respect to the bigrading. We put

 \be
 d=d_0 + d_1 + \t_0 + \t_1
 \la{3.3}
 \ee

where $d_0 (d_1)$ is the even (odd) derivative with bidegrees
$(1,0)$ and $(0,1)$ respectively, while $\t_0$ and $\t_1$ have
bidegrees $(-1,2)$ and $(2,-1)$. These two latter operators are
purely algebraic and involve the dimension-zero and
dimension-three-halves components of the torsion tensor
respectively. The subscripts indicate that $\t_0$ is an even
operator while $\t_1$ is odd. Explicitly, for $\o\in \O^{p,q}$,

 \bea
 (d_0 \o)_{a_1\ldots a_{p+1}\b_1\ldots \b_q}&=&
 \nab_{[a_1} \o_{a_2\ldots a_{p+1}]\b_1 \ldots \b_q}
+{p\over2}T_{[a_1 a_2}{}^c \o_{|c|a_3\ldots a_{p+1}]\b_1\ldots
\b_q} \nn\\
&\phantom{=}&+ q(-1)^pT_{[a_1(\b_1}{}^{\c}\o_{a_2\ldots
a_{p+1}]|\c|\b_2\ldots \b_q)}\\
&&\nn\\
(d_1 \o)_{a_1\ldots a_{p}\b_1\ldots \b_{q+1}}&=&
 (-1)^p\nab_{(\b_1} \o_{a_1\ldots a_{p}\b_2 \ldots \b_{q+1})}
+{q\over2}T_{(\b_1 \b_2}{}^{\c} \o_{a_1\ldots a_{p}]|\c|\b_3\ldots
\b_{q+1})} \nn\\
&\phantom{=}&+ p(-1)^p  T_{(\b_1[a_1}{}^{c}\o_{|c|a_2\ldots
a_{p}]\b_2\ldots \b_{q+1})}\\
&&\nn\\
(\t_0\o)_{a_1\ldots a_{p-1}\b_1\ldots \b_{q+2}}&=&
{p\over2}T_{(\b_1\b_2}{}^c\o_{|c|a_1\ldots a_{p-1}\b_3\ldots
\b_{q+2})}\\
&&\nn\\
(\t_1\o)_{a_1\ldots a_{p+2}\b_1\ldots \b_{q-1}}&=&
{q\over2}T_{[a_1 a_2}{}^{\c}\o_{a_3\ldots a_{p+2}]|\c|\b_1\ldots
\b_{q-1})}
 \eea

 The fact that $d^2=0$ implies the following identities:

 \bea
 \t_0^2 &=& 0
 \la{3.4}\\
 d_1\t_0+ \t_0 d_1&=&0
 \la{3.5}\\
 d_1^2 + d_0\t_0 + \t_0 d_0&=&0
 \la{3.6}\\
 d_0 d_1 + d_1 d_0 + \t_0\t_1+ \t_1\t_0&=&0
 \la{3.7}\\
 d_0^2 +d_1\t_1 + \t_1 d_1 &=&0
 \la{3.7.1}\\
 d_0\t_1 + \t_1 d_0 &=& 0
 \la{3.8}\\
 \t_1^2 &=&0
 \la{3.9}
 \eea

It should be noted that, although $d^2=0$ is an identity by the
definition of $d$, when one introduces a splitting of the tangent
space into even and odd, the equations  above (except for the
first and last ones) are only identities provided that the torsion
tensor obeys the first Bianchi identity. For example, if one
applies \eq{3.5} to an even one-form one finds that it is true
provided that $I_{\a\b\c}{}^d=0$.

Equation \eq{3.4} implies that we can consider the cohomology of
$\t_0$, as first noted in \cite{Bonora:mt}. We set

 \be
 H^{p,q}_{\t}=\{\o\in \O^{p,q}|\t_0\o=0\ {\rm mod}\ \o=\t_0 \l,
 \l\in \O^{p+1,q-2}\}
 \la{3.10}
 \ee

We can now define a spinorial derivative $d_F$ which will act on
elements of $H^{p,q}_{\t}$. If $\o\in[\o]\in H^{p,q}_{\t}$ we set

 \be
 d_F[\o]:=[d_1 \o]
 \la{3.11}
 \ee

It is easy to check that this is well-defined, i.e. $d_1\o$ is
$\t_0$-closed, and $d_F [\o]$ is independent of the choice of
representative. Furthermore it is simple to check that $d_F^2=0$.
This means that we can define the spinorial cohomology groups
$H^{p,q}_F$ in the obvious fashion.

If the dimension zero torsion tensor is flat, i.e. equal to the
gamma matrix, then the cohomology groups $H_F^{0,q}$ are
isomorphic to Berkovits's B-pure spinor cohomology groups and the
spinorial cohomology groups introduced in \cite{cntc}. The present
definition extends this to arbitrary dimension-zero torsion and to
mixed forms.

The above construction can be extended to vector-valued forms of a
certain type. Let $\O^p\otimes T$ denote the space of
vector-valued $p$-forms on a manifold. Any  $h\in \O^p\otimes T$
defines a derivation of degree $p-1$ of the algebra of
differential forms, $\O^q\ni \o\mapsto i_h\o\in \O^{p+q-1}$.
Essentially, the vector index of $h$ is contracted with one of the
indices of $\o$ while the remaining indices of $h$ and $\o$ are
antisymmetrised. We can combine this operation with exterior
differentiation to get a generalisation of the notion of Lie
derivative:

 \be
 \cL_h w:=d i_h \o + (-1)^p i_h d\o
 \la{3.12}
 \ee

Now let $h \in\O^{0,p}\otimes B$, the space of $(0,p)$-forms
taking their values in the even tangent space $B$, and let $\o\in
\O^{1,0}$. If we restrict the derivative to act in the odd
direction we get an expression which is tensorial in $\o$,

 \be
 d_1 h (\o):=d_1 i_h \o + (-1)^p i_h d_1\o
 \la{3.13}
 \ee

It is easy to check that this has the desired tensorial property
that $d_1 h( f\o)=f d_1 h(\o)$. In order to define a spinorial
cohomology, we have to remove the unwanted parts of $h$. We can
define an action of the dimension-zero torsion  in two ways,
either by converting an even form index into two odd ones as for
ordinary forms, or by converting an odd vector index into an even
vector index and an odd form index. In components,

 \be
 k_{a \b_1 \ldots \b_{p-2} }{}^b\mapsto
 T_{(\b_1\b_2}{}^c k_{c \b_3 \ldots \b_{p}) }{}^b
 \la{3.14}
 \ee

or

 \be
 k_{\a_1\ldots  \a_{p-1}}{}^{\c}\mapsto
 k_{(\a_1 \ldots \a_{p-1}}{}^{\c} T_{\c \a_p)}{}^c
 \la{3.15}
 \ee

For the case that $T_{\a\b}{}^c$ is covariantly constant one can
show that $d_1$ defines a derivative $d_F$ which squares to zero
on the spaces $\O^{0,p}\otimes B$ modulo the two equivalences
defined above. We shall refer to this as vector-valued spinorial
cohomology and denote it by $\check H^p_F$. In the case when one
has a general dimension-zero torsion one also has to factor out
terms of the form $k_{(\a_1\ldots\a_{p-2}}{}^D \nab_D
T_{\a_{p-1}\a_p)}{}^c$.

An application of this is to the supergravity equations of motion.
We saw above that these are implied by the constraint
$T_{\a\b}{}^c=-i(\c^c)_{\a\b}$. If we now make a deformation of
the supervielbein of the form $E_A{}^M \d E_M{}^B=h_A{}^B$ we find
that the change in the dimension zero torsion is

 \be
 \d T_{\a\b}{}^c =2\nab_{(\a} h_{\b)}{}^c + T_{\a\b}{}^D h_D {}^c
 -2h_{(\a}{}^D T_{D \b)}{}^c
 \la{3.16}
 \ee

The equations of motion will be satisfied for the deformed
supervielbein if this expression vanishes. Now the essential
deformation is given by $h_{\a}{}^b$ since the other components of
$h_A{}^B$ which appear in \eq{3.16} are simply field
redefinitions. So the esential part of \eq{3.16} is obtained by
ignoring the terms of the form $T_{\a\b}{}^d h_d{}^c$ and
$h_{(\a}{}^{\d} T_{\d \b)}{}^c$ which are precisely the
equivalences defined in \eq{3.14} and \eq{3.15}. Moreover,
changing $h_{\a}{}^c$ by a term of the form $k^{\c} T_{\c\a}{}^c$
also leads to a field redefinition. Therefore we see that the
possible deformations of the supervielbein which preserve the
field equations are given by $\check H^1_F$. This space therefore
tells us about on-shell degrees of freedom of the theory.


\subsection*{Deformations in the geometrical sector}

We shall now consider deformations of the supergravity equations
of motion. By this we mean we look for solutions of the Bianchi
identities in a power series in some dimensionful parameter, $t$
say, with the deformations being expressed in terms of the lowest
order fields. Such a solution, if consistent, will result in
modified equations of motion. The lowest-order fields are
themselves taken to be independent of $t$ although if we solve the
equations of motion perturbatively they will depend on the
parameter as well as on the solutions to the original supergravity
equations in a non-local manner.

This has been discussed for $D=11$ supergravity in \cite{cgnn}. We
begin by looking at this problem from the geometrical point of
view. As observed in \cite{cntc} one can look at the first-order
deformations in terms of the vector-valued cohomology group
$\check H^2_F(phys)$, where the notation indicates that in this
case the coefficients are not freely given superfields but are
instead functions of the physical fields. This is easily seen. The
dimension one-half Bianchi identity is

 \be
 \nab_{(\a} T_{\b\c}{}^d + T_{(\a\b}{}^E T_{|E|\c)}{}^d=0
 \la{3.17}
 \ee

If we expand in powers of $t$ in the form $T=\stackrel{0}{T} + t
\stackrel{1} T + \ldots$ we have at first order

 \be
 \nab_{(\a} \Ti_{\b\c)}{}^d + \To_{(\a\b}{}^e \Ti_{|e|\c)}{}^d
 + \Ti_{(\a\b}{}^{\e} \To_{|\e|\c)}{}^d=0
 \la{3.18}
 \ee

We see that the two quadratic terms here are of the form of
equivalences (with respect to the lowest order dimension zero
torsion) while the algebraic field redefinitions of
$\Ti_{\a\b}{}^c$ show that this equation can be interpreted as
$d_F[\tai_0]=0$. \footnote{We use the notation $\t_0$ to denote
the dimension zero part of the torsion considered as a
vector-valued two-form. $\tai_0$ denotes the first-order
deformation of this form.} Furthermore, one can make field
redefinitions of the supervielbein which will change
$[\tai_{\small{0}}]$ by the $d_F$ of a vector-valued one-form.
Hence the deformations of the equations of motion are indeed
characterised by $\check H^2_F(phys)$. It might be the case that
there are further constraints arising at higher order, although
this seems to be unlikely. We have seen that the theory can in
principle be discussed in a framework where we do not need to
introduce a connection or curvature; when we do, we expect to find
that the geometrical Bianchi identities should serve merely to
solve for the higher-dimensional components of the torsion and the
curvature in terms of the components of ${\Ti}_{\a\b}{}^c$, these
components being themselves given  in terms of the basic massless
supergravity field strength tensors.

{}From a practical point of view, however, the purely geometrical
approach is not likely to be very helpful in the task of finding
the explicit form of the deformation. It is expected that the
first deformation will be associated with an $R^4$ Lagrangian, so
that the parameter $t$ would be of the form $\ell^6$ where $\ell$
is some length scale. The analysis of the cohomology in this
situation would be extremely difficult to carry out.

For this reason it seems that it may be advantageous to introduce
the four-form field strength from the beginning. In particular, it
is easier to analyse the question of higher-order corrections, and
the cohomology groups which arise are based on ordinary
differential forms and not on vector-valued ones.

\section{Four-form supergravity}

Let us now study the theory with the inclusion of a four-form
field strength $G_4=dC$ from the beginning. The four
lowest-dimensional Bianchi identities $I=dG_4=0$ read

 \bea
 d_1 G_{0,4} + \t_0 G_{1,3}&=& 0\\
 \la{4.1}
 d_0 G_{0,4} + d_1 G_{1,3} + \t_0 G_{2,2}&=& 0\\
 \la{4.2}
 d_0 G_{1,3} + d_1 G_{2,2} + \t_0 G_{3,1} + \t_1 G_{0,4}&=& 0\\
 \la{4.3}
 d_0 G_{2,2} + d_1 G_{3,1} + \t_0 G_{4,0} + \t_1 G_{1,3}&=&0
 \la{4.4}
 \eea

We also note that $\t_0 G_{0,4}=0$. This is trivially true but
helps to make the picture clearer from the point of view of
cohomology.

Let us begin by assuming that $G_{0,4}=0$. Since

 \be
 G_{0,4}=d_1 C_{0,3} + \t_0 C_{1,2}
 \la{4.5}
 \ee

it is enough to take $[G_{0,4}]\in H^{0,4}_{\t}$ to be zero.
Moreover, since the gauge transformation of $C_{0,3}$ is

 \be
 \d C_{0,3} =d_1 Y_{0,2} + \t_0 Y_{1,1}
 \la{4.6}
 \ee

it follows that $[G_{0,4}]=0$ is equivalent to specifying an
element of $H^{0,3}_F$.

The $(0,5)$ component of the Bianchi identity then implies that

 \be
 \t_0 G_{1,3}=0
 \la{4.7}
 \ee

We can examine this equation perturbatively. Since $\Go_{1,3}=0$
we only need to use the lowest-order torsion. We therefore find

 \be
 (\c^a)_{(\a\b} \Gi_{a\c\d\e)}=0
 \la{4.8}
 \ee

which has the solution

 \be
 \Gi_{a\c\d\e}=(\c^b)_{(\c\d} K_{ab \e)} + (\c_{ab})_{(\c\d}
 L_{\e)}{}^b
 \la{4.9}
 \ee

The first term on the RHS is the trivial solution since
$K_{ab\c}\in \O^{2,1}$, while the second term is non-trivial
except for the purely spin one-half part of $L_{\e}^b$ which can
clearly be transferred to the first term. The trivial term can be
removed by a conventional constraint since $G_{1,3} = d_0 C_{0,3}
+ d_1 C_{1,2} + \t_0 C_{2,1}$. The second term can be removed by a
field redefinition of the supervielbein because

 \be
 \Go_{\a\b cd}=-i(\c_{cd})_{\a\b}
 \la{4.10}
 \ee

A (first-order) change in the supervielbein of the form $\d
E_{\a}=-h_{\a}{}^b E_b$ will therefore give rise to a change in
$G_{1,3}$ of precisely the same form as the second term in
\eq{4.9}. As we have seen, the gamma-trace part of this change
will not be required as one might expect from earlier remarks. So
we can take $\Gi_{1,3}=0$.

The $(1,4)$ component of the Bianchi identity then implies that

 \be
 \t_0 G_{2,2}=0
 \la{4.11}
 \ee

Writing this out in terms of perturbations we have

 \be
 (\c^b)_{(\a\b}\Gi_{ab \c\d)}+ \Ti_{(\a\b}{}^b (\c_{ab})_{\c\d)}=0
 \la{4.12}
 \ee

The only non-zero solutions to this equation can be removed by
field redefinitions. In particular, one can have solutions where
both tensors are scalar functions multiplied by the lowest-order
constant tensors. These can be set to zero by rescaling the even
and odd parts of the supervielbein separately.

At this stage we have shown that there are no non-trivial
deformations of the dimension-zero torsion and hence that one
would expect to recover the equations of motion of supergravity.
However, this is not quite right as we are using a Lorentzian
formalism and we have already used up the scale transformation to
deal with the dimension-zero component of $G$. We therefore have
to check at dimension one-half. Given that $I_{0,5}=I_{1,4}=0$ and
that $dI=0$ we find that the information contained in $I_{2,3}$
satisfies $\t_0 I_{2,3}=0$. This equation is solved by

 \be
 I_{ab\c\d\e}=(\c^c)_{(\c\d} J_{abc \e)} + (\c_{ab})_{(\c\d}
 K_{\e)}
 \la{4.13}
 \ee

where the first term is $\t_0$ trivial and the second is not;
indeed, the spin-half term in $J$ does not have the same
gamma-matrix structure as $K$. The equation itself is

 \be
 d_1 {\Go}_{2,2} + \t_0 {\Gi}_{3,1}=0
 \la{4.14}
 \ee

Note that the first term is non-zero because $d_1$ contains the
dimension one-half torsion. There is a spin-half field in the
first term (contained in the dimension one-half torsion) and a
spin-one-half field in $\Gi_{3,1}$; but, as we have just seen,
there are two independent constraints on these fields so that both
of them must vanish. It follows from \eq{4.13} that the remainder
of $\Gi_{3,1}$ vanishes too.

We therefore conclude  that any deformation of the supergravity
equations of motion for which $[G_{0,4}]=0$ is trivial. Having
established this to first order we can iterate the procedure. In
this perturbative sense, therefore, we can say that the equations
of motion of supergravity are determined by an element of
$H^{0,3}_F$. This seems to depend on a choice of an odd tangent
bundle, but it then turns out that this choice must be such that
the dimension-zero torsion is trivial.

\subsection*{Deformations in the four-form theory}

We now go on to consider the deformations of the supergravity
equations which arise when $[G_{0,4}]\neq 0$. The lowest-dimension
Bianchi identity $I_{0,5}$ reads

 \be
 d_1 G_{0,4}+ \t_0 G_{1,3}=0
 \la{4.15}
 \ee

Clearly this implies that $[d_1 G_{0,4}]=0$ thereby determining an
element of the cohomology group $H^{0,4}_F$. For a physical
deformation of the theory we require $G_{0,4}$ to be a function of
the supergravity fields, so we are really interested in elements
of $H^{0,4}_F(phys)$. We claim that, given such an element, the
rest of the Bianchi identities determine the other components of
$G_4$ in terms of derivatives of $G_{0,4}$ up to field
redefinitions and gauge transformations. In addition, the
dimension-zero torsion is determined, i.e the tensors $X$ and $Y$
at first order, as well as the dimension one-half spinor which
vanishes in the undeformed theory but which will be a function of
known fields in the deformed case. We shall give a brief outline
of this below and a more detailed account in terms of the
representations involved in section 6.

Perhaps the easiest way to see the result is to examine the
equation $dI_5=0$ for the components of the Bianchi identity. If
$I_{0,5}$ is satisfied then the information in $I_{1,4}$ will be
subject to the constraint $\t_0 I_{1,4}=0$ and so on, provided
that the geometrical Bianchi identities are satisfied.  However,
from dimension one-half onwards we need to take this into account
too. In this way we get a sequence of algebraic equations whose
solution will tell us the information contained in each $I_{p.q}$
given that the lower-dimensional ones have been solved.

Assuming that $I_{0,5}=0$ we find $\t_0 I_{1,4}=0$ which in
components is solved by

 \be
 I_{a\b\c\d\e}=(\c^b)_{(\b\c}J_{ab \d\e)}+
 (\c_{ab})_{(\b\c}K_{\d\e)}{}^b
 \la{4.16}
 \ee

>From this we can deduce that $I_{1,4}$ allows us to solve for the
dimension-zero torsion and $G_4$ deformations up to field
redefinitions as in the previous case. To go from $I_{1,4}$ to
$I_{2,3}$ we have to take into account the fact that $G_{2,2}$ has
a non-vanishing component at zeroth order, so that we are not
allowed to use the equation $d_1\t_0+\t_0 d_1=0$ immediately. One
finds, by differentiating $I_{1,4}$ and then setting the result to
zero that
 \be
 (d_1\t_0+\t_0 d_1)G_{2,2}= \t_0 I_{2,3}
 \la{4.17}
 \ee

or, in indices, and remembering that the Bianchi identities are
satisfied to zeroth order,

\be I_{(\a\b\c}{}^b (\c_{ab})_{\d\e)}= (\c^b)_{(\a\b}
I_{ab\c\d\e)} \la{4.18}\ee

This equation is solved by

\be I_{ab\c\d\e}=(\c^c)_{(\c\d} J_{abc\e)} +(\c_{ab})_{(\c\d}
K_{\e)} +2(\c_{c[a})_{(\c\d} L_{\e) b}{}^c
-M_{(\c\d}{}^{\h}(\c_{ab})_{\e)\h}\la{4.18.1}\ee

and

\be I_{\a\b\c}{}^b= (\c^c)_{(\a\b} L_{\c) c}{}^b + M_{(\a\b}{}^\h
(\c^b)_{\c)\h} \la{4.18.2}\ee

Furthermore, the components of $L$ and $M$ which correspond to
field redefinitions of the dimension one-half connection and the
choice of even tangent bundle cancel out in these expressions.
Setting $L$ and $M$ equal to zero therefore determines the
dimension one-half torsion, up to these field redefinitions, while
setting $J=0$ determines the dimension one-half component of
$G_4$. Finally, if we put $K=0$ we determine the dimension
one-half Weyl fermion as an explicit function of the fields at
first order. The remainder of the dimension one-half geometrical
Bianchi identity must be automatically satisfied by the deformed
dimension zero torsion determined from $G_{0,4}$ up to
equivalences of the type given in \eq{3.14} and \eq{3.15}, and
this therefore implies that the equation $d_F[\tai_0]=0$ discussed
earlier is satisfied.

We can continue this analysis at the next order where we encounter
the dimension one geometrical Bianchi identities. However, it is
likely that they are automatically satisfied if the dimension
one-half one is and so, at this stage, we can assume that the
d-algebra is satisfied. We then find that $\t_0 I_{3,2}=0$ which
allows us to relate $G_{4,0}$ to the dimension-one torsion. To
summarise this section, we have shown that the first-order
deformation of the theory is completely determined by an element
of $H^{0,4}_F(phys)$ given which one can compute the first-order
deformations of all of the remaining components of the four-form
field strength as well as the components of the torsion and
curvature in a systematic way using the Bianchi identities.


\section{Seven-form supergravity}

It is well-known that one can  introduce a seven-form field
strength $G_7$ as well as the four-form which we shall now denote
by $G_4$. In supergravity itself it obeys the Bianchi identity

 \be
 dG_7=\half G_4^2
 \la{5.1}
 \ee

The only non-vanishing components of $G_7$ are $G_{7,0}$ which is
the dual of $G_{4,0}$ and $G_{5,2}$ which is given by

 \be
 G_{abcde \a\b}=-i(\c_{abcde})_{\a\b}
 \la{5.2}
 \ee

This formulation was first discussed in superspace in
\cite{Candiello:di}.

In the deformed theory, however, we have to take into account the
Chern-Simons term which is necessary in order to cancel the
anomaly on the fivebrane \cite{Duff:1995wd}. The modified Bianchi
identity is

 \be
 dG_7=\half G_4^2 + \b X_8
 \la{5.3}
 \ee

where $\b$ is a parameter of dimension $\ell^6$ and
$X_8=\tr(R^4)-{1\over4}(\tr(R^2))^2$. The necessity of including
$X_8$ in the superspace Bianchi identity was observed in
\cite{cgnn}, although it should be noted that $G_7$ is
not the superspace dual of $G_4$, and neither is $G_{7,0}$ the
dual of $G_{4,0}$ in the presence of higher-order corrections. In
the general  \eq{5.3} will determine $G_{7,0}$ to be the dual of
$G_{4,0}$ plus many other correction terms with one or more powers
of $\b$.

Provided that this equation is consistent, so that one can find a
particular solution corresponding to the $X$-term (which can be
constructed from the zeroth order supergravity fields), the
question of uniqueness then boils down to solving the original
identity, \eq{5.1}, at first order in $\b$. If $G_{0,7}\neq 0$,
then it defines an element of $H^{0,7}_F(phys)$. If this group is
trivial (at this dimension) then we must consider $G_{1,6}$ which
will satisfy $\t_0 G_{1,6}=0$ in this case. However, the group
$H^{1,6}_{\t}$ is not trivial, and  this will lead to an element
of the spinorial cohomology group $H^{1,6}_F(phys)$. If this is
trivial as well at this dimension, we shall have $\t_0 G_{2,5}=0$.
The cohomology group $H^{2,5}_{\t}$ is also non-trivial and so we
get an element of $H^{2,5}_F(phys)$.  If this element should also
vanish, the next Bianchi, at dimension $-1$ involves the $G_4$
field, and reads, given that $G_{0,7}=G_{1,6}=G_{2,5}=0$,

 \be
 \tao_0 \Gi_{3,4}=\Go_{2,2} \Gi_{0,4}
 \la{5.4}
 \ee

In index notation,

 \be
 (\c^c)_{(\a\b} \Gi_{abc \c\d\e\h)}=(\c_{ab})_{(\a\b}\Gi_{\c\d\e\h)}
 \la{5.5}
 \ee

This equation admits a trivial solution for $\Gi_{3,4}$ and the
$\t_0$-trivial expression for $\Gi_{0,4}$ can also easily be seen
to be a solution for a simple choice of the left-hand side. But
there are no other solutions, and this implies that both $G_{3,4}$
and $G_{0,4}$ are unchanged at first order.  To see this one can
multiply \eq{5.5} by six factors of a B-pure spinor $u$. The
left-hand side is annihilated due to the $\c^c$ factor, but the
right-hand side is not. Indeed, only the $\t_0$-trivial
representation in $G_{0,4}$ is killed. There are three non-trivial
representations in $G_{0,4}$ which are $(n,k)$-tensors of the form
$(2,2)$, $(2,5)$ and $(5,5)$. None of these is annihilated so that
\eq{5.5} indeed implies that $G_{0,4}$ is  $\t_0$-trivial. The
triviality of $G_{3,4}$ then folllows from the fact that
$H^{3,4}_{\t}=0$. From the earlier discussion of the $G_4$ Bianchi
identity, it is clear, at this stage, that we recover the standard
zeroth order theory.

This analysis shows that the possible ambiguities in solving the
Bianchi identities at order $\b$ are described by the cohomology
groups $H^{0,7}_F(phys),\,H^{1,6}_F(phys)$ and $H^{2,5}_F(phys)$.
Moreover, a solution of the modified Bianchi identity will
automatically determine the first-order correction to $G_{0,4}$
and hence the modified equations of motion. Therefore, we conclude
that, if these three cohomology groups are zero at first order in
$\b$, the complete equations of motion and supersymmetry
transformations can be determined systematically by solving the
Bianchi identities.

We shall now argue that it is indeed the case that these groups
vanish at this order, if we assume that the invariants we should
consider should be at least quartic in the fields.  In fact, if
this assumption is made, there is nothing one can write down for
$G_{0,7}$ of the right dimension.  However, for $G_{1,6}$ one
could have terms of the form $W^4$, while for $G_{2,5}$ we could
have terms of the form $W^3\Psi$ where $\Psi$ is the gravitino
field strength. We shall argue below by an indirect argument that
this is not possible, and so we conclude that we can write
$X_{0,8}$ in the form $\t_0 G_{1,6}$ and that this solves for
$G_{1,6}$ up to field redefinitions.

\subsection*{Action principle}

One way of going about the construction of integral
super-invariants in $D$ dimensions is to consider a closed
$D$-form $L_D$, say, up to exact terms which would modify the
integrand by a total derivative
\cite{D'Auria:1982pm,Gates:1997ag}. The invariant is given by

 \be
 S=\int\, d^Dx\, \e^{m_1\ldots m_D} L_{m_1\ldots m_D}|
 \la{5.6}
 \ee

where the bar denotes the evaluation of a superfield at $\th=0$.

This construction yields an expression which is automatically
invariant under local supersymmetry transformations and spacetime
diffeomorphisms. Under a infinitesimal diffeomorphism in
superspace one has

 \be
 \d L_D = \cL_v L_D= di_v L_D + i_v dL_D=d i_v L_D
 \la{5.7}
 \ee

from which one reads that the integrand above transforms as a
total derivative under the spacetime transformations corresponding
to zeroth order terms in the superspace vector field $v$. These
are just local supersymmetry transformations and spacetime
diffeomorphisms.

There is a class of actions, namely those terms of the
Chern-Simons type, which can be usefully discussed using this
method. In this case there is a closed $D+1$-form, $W_{D+1}$,
which can be written in two ways: as the exterior derivative of
the potential $Z_D$, the Chern-Simons term,
 which defines $W$ or as
the exterior derivative of a well-defined $D$-form $K_D$, say. The
difference between the two then automatically gives a closed
Lagrangian form $L_D=K_D-Z_D$. The potential term obviously
contains the Chern-Simons term in the Lagrangian. This method can
be used to derive in a systematic way the Green-Schwarz actions
for branes in the superembedding formalism starting from the
Wess-Zumino term. \cite{Bandos:1995dw,Howe:1998ts}.

In the present case, the twelve-form we need to consider is

 \be
 W_{12}= \half G_4^3 + 3\b G_4 X_8
 \la{5.8}
 \ee

In fact, this form should lead to the sum of the zeroth- and
first-order actions provided one exercises sufficient care with
the use of an on-shell formalism.

Now suppose this does not give a unique invariant of $R^4$ type.
In that case we would expect that there should be a closed
$11$-form, $L_{11}$. However, we claim that the cohomology class
of such a form must be trivial to order $\b$ provided that we
restrict ourselves to terms which are at least quartic in the
fields. To see this we note that the dimension of $L_{11,0}$ is 8
and since $W$ has dimension 1, the lowest-dimensional object
available is $W^4$ at dimension 4. This corresponds to the
component $L_{3,8}$ and so we must infer that
$L_{2,9}=L_{1,10}=L_{0,11}=0$. But then we must have $\t_0
L_{3,8}=0$. However,  it seems likely that all of the
$\t$-cohomology groups $H^{p,q}_{\t}$ vanish for $p\geq 3$. This
can be supported by the principle of maximal propagation of
representations presented in \cite{cntc}, but one should be aware
that this does not give an unambiguous result. On the other hand,
all of the known non-trivial examples of $\t$-cohomology involve a
factor of $\c_{ab}$ and make use of the membrane identity

\be (\c^a)_{(\a\b} (\c_{ab})_{\c\d)}=0 \la{5.9} \ee

When $p\geq 3$ the putative cocycle must involve another index and
there is no obvious way of doing this without introducing a second
factor of $\c_2$. However, one then runs into the five-brane
identity

\be (\c^a)_{(\a\b} (\c_{abcde})_{\c\d)}=(\c_{[ab})_{(\a\b}
(\c_{cd]})_{\c\d)} \la{5.10} \ee

which implies a cocycle with two $\c_2$s is in fact a coboundary.
Assuming this to be correct, we conclude that any solution to $d
L_{11}=0$ at $\ell^6$ which is at least quartic in the fields must
be trivial.


\section{The $G_4$ Bianchi identities}

In this section we give a more detailed discussion of the
four-form Bianchi identity at dimension $-\frac{1}{2}$ and we
derive a number of constraints on $G_{0,4}$. We then proceed to
dimension zero to give the expression of the zero-dimension
torsion in terms of (derivatives of) the superfields in $G_{0,4}$.
We have found the programs LiE \cite{cohen} and Gamma
\cite{Gran:2001yh} useful for representation-theoretic
considerations and Gamma-matrix computations respectively. An
explanation of the group-theoretic notation is given in appendix
C.

To zeroth order the only nonzero components of $G$ are
$$
G_{abcd}(=W_{abcd}); \qquad
G_{ab\alpha\beta}=-i(\gamma_{ab})_{\alpha\beta}
$$
In components the BI reads,
\begin{equation}
\nab_{[A_1}G_{A_2\dots A_5\}}+2T_{[A_1A_2|}{}^F
G_{F|A_3A_4A_5\}}=0
\end{equation}
The invariance $G\rightarrow G+dC$ reads in components,
\be \d G_{A_1\dots A_4} =\frac{3}{2}T_{[A_1A_2|}{}^FC_{F|A_3A_4\}}
+\nab_{[A_1}C_{A_2A_3 A_4\}} \label{redone}. \ee
In addition we have the freedom of redefinitions of the vielbein
$E_{A}{}^M\delta E_M{}^B:=h_A{}^B$. These induce shifts in $G$
according to,
\begin{equation}
\delta G_{A_1\dots A_4}=-4h_{[A_1|}{}^FG_{F|A_2A_3A_4\}}
\label{redtwo}
\end{equation}
Note that we are not free to vary all the components of $E_M{}^A$,
because some of them have already been used to remove some
components of the torsion. The only components of the vielbein
which are still free to use in redefinitions of the type above,
are \cite{cgnn},
$$
h_{\alpha}{}^a\vert_{(00001)}; \qquad
h_{\alpha}{}^a\vert_{(10001)}; \qquad
h_{\alpha}{}^\beta\vert_{(00000)}; \qquad
h_{\alpha}{}^\beta\vert_{(01000)}
$$
\subsection*{The BI at dimension $-\frac{1}{2}$.}
\noindent To first order the dimension $-\frac{1}{2}$ Bianchi
identity reads,
\begin{equation}
0=\nab_{(\alpha_1} G_{\alpha_2\dots \alpha_5)}
-2i(\gamma^f)_{(\alpha_1\alpha_2|}
G_{f|\alpha_3\alpha_4\alpha_5)}.
\end{equation}
In order to solve it we need the expansions, in irreducible
representations, of the fields
 $G_{\alpha_1\dots \alpha_4}$,
$\nab_{\alpha_1}G_{\alpha_2\dots \alpha_5}$ and
$G_{a\alpha_1\alpha_2\alpha_3}$. The field $G_{\alpha_1\dots
\alpha_4}$ decomposes as,
\begin{alignat}{2}
\otimes^4_s(00001) \sim &(00002) \oplus (00010) \oplus (00100)
\oplus (00000) \oplus (00004)
\oplus (01002)\nn\\
&\oplus (10002) \oplus (02000) \oplus (11000) \oplus (20000)\nn
\end{alignat}
Redefinitions $\delta G_{\alpha_1\dots \alpha_4}$ of the type
(\ref{redtwo}) vanish at this order. Redefinition (\ref{redone})
reads
\begin{equation}
\delta G_{\alpha_1\dots
\alpha_4}=-\frac{3i}{2}(\gamma^f)_{(\alpha_1\alpha_2|}
C_{f|\alpha_3\alpha_4)}+\nab_{(\alpha_1}
C_{\alpha_2\alpha_3\alpha_4)}
\end{equation}
The superfield $C_{a\alpha_1\alpha_2}$ decomposes as
\begin{alignat}{2}
\otimes_s^2(00001)\otimes(10000)\sim
&(00002)\oplus(00010)\oplus(00100)\oplus(01000)\oplus(10000)\nn\\
&\oplus(00000) \oplus(10002)\oplus(11000)\oplus(20000)\nn
\end{alignat}
Therefore, we can use all but the (01000) and (10000) irreps in
$C_{a\alpha_1\alpha_2}$ to fix,
\begin{equation}
G_{\alpha_1\dots \alpha_4} \sim \oplus (00004) \oplus (01002)
\oplus (02000)
\end{equation}
The superfield $C_{\alpha_1\alpha_2\alpha_3}$ will not be used at
this stage.
Explicitly we expand,
\begin{alignat}{3}
G_{\alpha_1\dots\alpha_4} &={1\over
8}(\gamma^{a_1a_2})_{(\alpha_1\alpha_2}
(\gamma^{b_1b_2})_{\alpha_3\alpha_4)}
A_{a_1a_2;b_1b_2}  &\qquad (02000) \nn\\
&+{1\over 240}(\gamma^{a_1\dots a_5})_{(\alpha_1\alpha_2}
(\gamma^{b_1b_2})_{\alpha_3\alpha_4)}
B_{a_1\dots a_5;b_1b_2}  &(01002) \nn\\
&+{1\over 28800}(\gamma^{a_1\dots a_5})_{(\alpha_1\alpha_2}
(\gamma^{b_1\dots b_5})_{\alpha_3\alpha_4)} C_{a_1\dots
a_5;b_1\dots b_5}  &(00004)
\end{alignat}
The field $G_{\alpha_1\dots\alpha_4}$ enters the BI through its
supercovariant derivative, so we also need to expand to level
$\theta^1$ in the fields $A,B,C$.

\noindent $D A_{a_1a_2;b_1b_2}$:
\begin{alignat}{2}
(00001)\otimes(02000)\sim (01001)\oplus(11001)\oplus(02001)\nn
\end{alignat}
Explicitly\footnote{To simplify the notation we will omit the
antisymmetrization brackets. Antisymmetrization is understood in
the $(a_1a_2)$ and $(b_1,b_2)$ indices. Similarly in (72,73)
below. A more detailed discussion of the projections
$\hat{\Pi}^{(p,q)}$ onto the irreducible part of a $(p,q)$-tensor
can be found in appendix A. The explicit form of the expansions
(71--73) is given in appendix B.}:
\begin{alignat}{3}
DA_{a_1a_2;b_1b_2}=&\hat{\Pi}^{(2,2)}(A^1_{a_1a_2;b_1b_2})
&\qquad(02001)\nn\\
+&\hat{\Pi}^{(2,2)}(\gamma_{b_1}A^1_{a_1a_2;b_2})&\qquad(11001)\nn\\
+&\hat{\Pi}^{(2,2)}(\gamma_{b_1b_2}A^1_{a_1a_2})&\qquad(01001)
\label{dia}
\end{alignat}
$DB_{a_1\dots a_5;b_1b_2}$:
\begin{alignat}{2}
(00001)\otimes(01002)\sim
&(00003)\oplus (00011)\oplus(00101)\oplus(01001)\nn\\
&\oplus(10003)\oplus(11001)\oplus (01003)\oplus (02001)\oplus
\dots\nn
\end{alignat}
where the ellipses stand for the irreps that do not enter the BI.
Explicitly:
\begin{alignat}{3}
DB_{a_1\dots a_5;b_1b_2}=&\hat{\Pi}^{(5,2)}
(\gamma_{b_1b_2}B^1_{a_1\dots a_5})&\qquad(00003)\nn\\
+&\hat{\Pi}^{(5,2)}
(\gamma_{b_1b_2a_1}B^1_{a_2\dots a_5})&\qquad(00011)\nn\\
+&\hat{\Pi}^{(5,2)}
(\gamma_{b_1b_2a_1a_2}B^1_{a_3a_4a_5})&\qquad(00101)\nn\\
+&\hat{\Pi}^{(5,2)}
(\gamma_{b_1b_2a_1a_2a_3}B^1_{a_4a_5})&\qquad(01001)\nn\\
+&\hat{\Pi}^{(5,2)}
(B^1_{a_1\dots a_5;b_1b_2})&\qquad(01003)\nn\\
+&\hat{\Pi}^{(5,2)}
(\gamma_{b_1}B^1_{a_1\dots a_5;b_2})&\qquad(10003)\nn\\
+&\hat{\Pi}^{(5,2)}
(\gamma_{a_1a_2a_3}B^1_{a_4a_5;b_1b_2})&\qquad(02001)\nn\\
+&\hat{\Pi}^{(5,2)}
(\gamma_{b_1a_1a_2a_3}B^1_{a_4a_5;b_2})&\qquad(11001)\nn\\
+&\dots & \label{dibi}
\end{alignat}
$DC_{a_1\dots a_5;b_1\dots b_5}$:
\begin{alignat}{2}
(00001)\otimes(00004)\sim (00003)\oplus(10003)\oplus(01003)\oplus
(00005)\oplus \dots\nn
\end{alignat}
Explicitly:
\begin{alignat}{3}
DC_{a_1\dots a_5;b_1\dots b_5}=&
\hat{\Pi}^{(5,5)}(\gamma_{a_1\dots a_5}C^1_{b_1\dots b_5})
&\qquad(00003)\nn\\
+&\hat{\Pi}^{(5,5)}(C^1_{a_1\dots a_5;b_1\dots b_5})&\qquad(00005)\nn\\
+&\hat{\Pi}^{(5,5)}(\gamma_{b_1b_2b_3}C^1_{a_1\dots a_5;b_4b_5})
&\qquad(01003)\nn\\
+&\hat{\Pi}^{(5,5)}(\gamma_{b_1\dots b_4}C^1_{a_1\dots a_5;b_5})
&\qquad(10003)\nn\\
+&\dots & \label{dici}
\end{alignat}
The field $G_{a \alpha_1\alpha_2\alpha_3}$ decomposes as,
\begin{alignat}{2}
\otimes^3_s(00001)\otimes(10000) \sim &(00003) \oplus (00011)
\oplus (00101)
\oplus 2(01001)\nn\\
&\oplus 3(10001) \oplus 2(00001) \oplus (10003) \oplus (11001)
\oplus (20001)\nn
\end{alignat}
Redefinition (\ref{redone}) reads
\begin{alignat}{2}
\delta G_{a\alpha_1\alpha_2 \alpha_3}&=
\frac{3}{4}T_{a(\alpha_1|}{}^\beta C_{\beta|\alpha_2\alpha_3)}
+\frac{1}{4}D_{a}C_{\alpha_1\alpha_2\alpha_3}
-\frac{3}{4}D_{(\alpha_1|}C_{a|\alpha_2\alpha_3)}\nn\\
&-\frac{3i}{4}(\gamma^f)_{(\alpha_1\alpha_2|}C_{fa|\alpha_3)}
\end{alignat}
The field $C_{a_1a_2\alpha}$ decomposes as,
\begin{equation}
(01000)\otimes(00001)\sim (01001)\oplus(10001)\oplus(00001)
\end{equation}
The irreps on the rhs above can be used to eliminate the
corresponding irreps in the decomposition of $G_{a\alpha_1\alpha_2
\alpha_3}$. The fields $C_{\alpha_1\alpha_2\alpha_3}$ and
$C_{a\alpha_1\alpha_2}$ will not be used at this stage.
Redefinitions $\delta G_{a\alpha_1\alpha_2 \alpha_3}$ of the type
(\ref{redtwo}) read,
\begin{alignat}{2}
\delta G_{a\alpha_1\alpha_2 \alpha_3}&=
3h_{(\alpha_1|}{}^fG_{fa|\alpha_2\alpha_3)}\nn\\
&=3i(\gamma_a{}^m)_{(\alpha_1\alpha_2|}V_{m|\alpha_3)}
+3i(\gamma_a{}^m)_{(\alpha_1\alpha_2|}(\gamma_mV)_{|\alpha_3)}
\label{blirp}
\end{alignat}
where we have decomposed the vielbein variation as
\begin{equation}
h_\alpha{}^a=(\gamma^a)_\alpha{}^\beta V_\beta+V^a_\alpha,
\end{equation}
into a vector-spinor $V_{a\alpha}$ and a spinor $V_\alpha$ part.
The vector-spinor part of the vielbein can be used to eliminate
one of the two remaining vector-spinors in the decomposition of
$G_{a\alpha_1\alpha_2 \alpha_3}$. Naively one might expect that
the spinor part of the vielbein could also be used in order to
eliminate the remaining spinor in $G_{a\alpha_1\alpha_2
\alpha_3}$. However this is not the case. The reason is that the
spinor part of the vielbein induces a redefinition (cf
(\ref{blirp}) above) of the same form as the redefinition induced
by the spinor part of $C_{ab\alpha}$,
\begin{equation}
\delta G_{a\alpha_1\alpha_2 \alpha_3}\vert_{(00001)} \sim
(\gamma^m)_{(\alpha_1\alpha_2|}(\gamma_{ma}C)_{|\alpha_3)}
\end{equation}
as can be seen using the identity
\begin{equation}
(\gamma^m)_{(\alpha_1\alpha_2|}(\gamma_{ma})_{|\alpha_3)}{}^\alpha=
-(\gamma_a{}^m)_{(\alpha_1\alpha_2|}(\gamma_{m})_{|\alpha_3)}{}^\alpha
\end{equation}
To conclude, after taking
 the field redefinitions above into account,
the expansion of $G_{a\alpha_1\alpha_2 \alpha_3}$ reads,
\begin{alignat}{2}
G_{a\alpha_1\alpha_2 \alpha_3} \sim &(00003) \oplus (00011) \oplus
(00101)
\oplus (01001)\nn\\
&\oplus (10001) \oplus (00001) \oplus (10003) \oplus (11001)
\oplus (20001)
\end{alignat}
We therefore arrive at the following explicit expansion,
\begin{alignat}{3}
i G_{a\alpha_1\alpha_2\alpha_3}&= {1\over
120}(\gamma_{a}{}^{i_1\dots i_5})_{(\alpha_1\alpha_2|}
(T_{i_1\dots i_5})_{|\alpha_3)}&
\qquad (00003) \nn\\
&+{1\over 24}(\gamma_{a}{}^{i_1\dots i_4})_{(\alpha_1\alpha_2|}
(T_{i_1\dots i_4})_{|\alpha_3)}
&(00011) \nn\\
&+{1\over 6}(\gamma^{i_1i_2})_{(\alpha_1\alpha_2|}
(T_{ai_1i_2})_{|\alpha_3)}
&(00101) \nn\\
&+{1\over 2}(\gamma^{i_1i_2})_{(\alpha_1\alpha_2|}
(\gamma_{a}T_{i_1i_2})_{|\alpha_3)}
&(01001) \nn\\
&+(\gamma^{i})_{(\alpha_1\alpha_2|} (\gamma_{i}T_{a})_{|\alpha_3)}
&(10001) \nn\\
&+(\gamma_{a})_{(\alpha_1\alpha_2|} (T)_{|\alpha_3)}
&(00001) \nn\\
&+{1\over 120}(\gamma^{i_1\dots i_5})_{(\alpha_1\alpha_2|}
(T_{i_1\dots i_5;a})_{|\alpha_3)}
&(10003) \nn\\
&+{1\over 2}(\gamma^{i_1i_2})_{(\alpha_1\alpha_2|}
(T_{i_1i_2;a})_{|\alpha_3)}
&(11001) \nn\\
&+(\gamma^{i})_{(\alpha_1\alpha_2|} (T_{i;a})_{|\alpha_3)}
&(20001)
\end{alignat}
The following table summarizes the various contributions to the BI
for each irrep:

\bigskip

\begin{center}

\begin{tabular}{|c||c|c|c|c|}\hline
B.I:& $G_{a\alpha_1\alpha_2\alpha_3}$:&$DC_{a_1\dots a_5;b_1\dots
b_5}$:& $DB_{a_1\dots a_5;b_1b_2}$:&$DA_{a_1a_2;b_1b_2}$:
\\\hline\hline
(00001)&1 &--&--&-- \\\hline (00003)&1 &1&1&--\\\hline (00005)&--
&1&--&--\\\hline (00011)& 1&--&1&--\\\hline (00101)&
1&--&1&--\\\hline (01001)& 1&--&1&1\\\hline (01003)&
--&1&1&--\\\hline (02001)& --&--&1&1\\\hline (10001)&
1&--&--&--\\\hline (10003)& 1&1&1&--\\\hline (11001)&
1&--&1&1\\\hline (20001)& 1&--&--&--\\\hline
\end{tabular}
\medskip
\end{center}

\noindent By inspection of the table above, we can guess that the
following will happen: a) The BI projected onto the irreducible
representations (00005), (01003) and (02001), will give three {\it
constraints} on the superfield $G_{\alpha_1\dots \alpha_4}$. b)
The (00001), (10001) and (20001) components of
$G_{a\alpha_1\alpha_2\alpha_3}$ will be set to zero by the BI. c)
The remaining  irreps of the BI will be used to express the rest
of the components of $G_{a\alpha_1\alpha_2\alpha_3}$ as linear
combinations of the components of $DG_{\alpha_1\dots \alpha_4}$.
Getting the precise relations  requires some tedious calculations.
Our strategy will be to plug in the BI the field expansions given
above and to solve by projecting onto each irreducible
representation using the formulae in appendices A. and B.

\noindent The solution to the dimension $-\frac{1}{2}$ BI reads,

\begin{alignat}{2}
T_{a_1\dots a_5} &=-\frac{11}{4}C_{a_1\dots a_5}^1
-2B_{a_1\dots a_5}^1\nn\\
T_{a_1\dots a_4}&=-\frac{26}{25}B_{a_1\dots a_4}^1\nn\\
T_{a_1a_2a_3}&=\frac{351}{175}B_{a_1a_2a_3}^1\nn\\
T_{a_1a_2}&=-\frac{9}{35}B_{a_1a_2}^1-A_{a_1a_2}^1\nn\\
T_{a}&=0\nn\\
T&=0\nn\\
T_{a_1\dots a_5;b}&=\frac{3}{5}C^1_{a_1\dots a_5;b}+\frac{1}{5}
B^1_{a_1\dots a_5;b}\nn\\
T_{a_1a_2;b}&=-\frac{63}{200}B^1_{a_1a_2;b}+\frac{1}{8}
A^1_{a_1a_2;b}\nn\\
T_{a;b}&=0.
\end{alignat}
The equations above are equivalent to the statement that the
superfields in $G_{1,3}$ are expressible in terms of the
superfields in $G_{0,4}$. In addition we have three constraints,
\begin{alignat}{2}
0=&C_{a_1\dots a_5;b_1\dots b_5}^1\nn\\
0=&B^1_{a_1\dots a_5;b_1b_2}-\frac{3}{10}C^1_{a_1\dots a_5;b_1b_2}\nn\\
0=&A^1_{a_1a_2;b_1b_2}-\frac{7}{5}B^1_{a_1a_2;b_1b_2}.
\end{alignat}

\subsection*{The BI at dimension 0.}
In order to solve for the dimension-zero torsion, it suffices to
focus on the (11000) and (10002) irreps, since these are the ones
that parametrise the deformations of ordinary supergravity. Let us
expand
\begin{equation}
T_{\alpha\beta}{}^a=-i(\gamma^a)_{\alpha\beta}+\frac{1}{2}
(\gamma^{e_1e_2})_{\alpha\beta}X_{e_1e_2;}{}^a+\frac{1}{120}
(\gamma^{e_1\dots e_5})_{\alpha\beta}X_{e_1\dots e_5;}{}^a
\end{equation}
and
\begin{equation}
G_{ab\alpha\beta}=-i(\gamma_{ab})_{\alpha\beta}
+S_{ab;e}(\gamma^e)_{\alpha\beta}+\frac{1}{60}S_{e_1\dots e_5;[a}
(\gamma_{b]}{}^{e_1\dots e_5})_{\alpha\beta}+\dots,
\end{equation}
where the ellipses stand for irreps in $G_{2,2}$ other than
$(11000)$ and $(10002)$. The BI at dimension 0 reads
\begin{alignat}{2}
0=&\frac{i}{6}\nab_a G_{\alpha_1 \dots \alpha_4}-\frac{2i}{3}
\nab_{(\alpha_1|}G_{a|\alpha_2\alpha_3\alpha_4)}+\frac{2i}{3}
T_{a(\alpha_1|}{}^\epsilon
G_{\epsilon|\alpha_2\alpha_3\alpha_4)}&\nn\\
&+(\gamma^f)_{(\alpha_1\alpha_2}
(\gamma^e)_{\alpha_3\alpha_4)}S_{fa;e}+\frac{1}{60}
(\gamma^f)_{(\alpha_1\alpha_2|}(\gamma_{[a|}{}^{e_1\dots
e_5})_{|\alpha_3\alpha_4)}
S_{e_1\dots e_5;|f]} &\nn\\
&+\frac{1}{2}(\gamma^{e_1e_2})_{(\alpha_1\alpha_2|}
(\gamma_{fa})_{|\alpha_3\alpha_4)}X_{e_1e_2;}{}^f+\frac{1}{120}
(\gamma^{e_1\dots
e_5})_{(\alpha_1\alpha_2|}(\gamma_{fa})_{|\alpha_3\alpha_4)}
X_{e_1\dots e_5;}{}^f& +\dots \label{zerodim}
\end{alignat}
Let us denote by $A^{(1)}_{a_1a_2;b},\, A^{(2)}_{a_1a_2;b}$ the
irreducible hook-tensor part of the contractions of the rhs of the
first line of (\ref{zerodim}) with
$$(\gamma^{ba})^{\alpha_1\alpha_2}
(\gamma^{a_1a_2})^{\alpha_3\alpha_4},$$
$$(\gamma^{b})^{\alpha_1\alpha_2}
(\gamma^{[a_1|})^{\alpha_3\alpha_4}\eta^{|a_2]a}$$ respectively.
We find,
\begin{alignat}{2}
X_{a_1a_2;b}&=\frac{-9A^{(1)}_{a_1a_2;b}+2A^{(2)}_{a_1a_2;b}  }
{2^{12} \times 5}&\nn\\
S_{a_1a_2;b}&=\frac{3A^{(1)}_{a_1a_2;b}-214A^{(2)}_{a_1a_2;b}  }
{2^{12}\times 3 \times 5 }&
\end{alignat}
Similarly, let us denote by $B^{(1,2,3)}_{a_1\dots a_5;b}$, the
irreducible hook-tensor part of the contractions of the rhs of the
first line of (\ref{zerodim}) with
$$
(\gamma^{ba})^{\alpha_1\alpha_2} (\gamma^{a_1\dots
a_5})^{\alpha_3\alpha_4},
$$
$$(\gamma^{e})^{\alpha_1\alpha_2}
(\gamma_e{}^{a_1\dots a_5})^{\alpha_3\alpha_4}\eta^{ab},$$
$$(\gamma^{b[a_1|})^{\alpha_1\alpha_2}
(\gamma^{|a_2\dots a_5]a})^{\alpha_3\alpha_4}$$ respectively. We
find,
\begin{alignat}{2}
X_{a_1\dots a_5;b}&=\frac{4B^{(1)}_{a_1\dots a_5;b}
-3B^{(2)}_{a_1\dots a_5;b}  }
{2^{13}}&\nn\\
S_{a_1\dots a_5;b}&=\frac{4B^{(1)}_{a_1\dots a_5;b}
-99B^{(2)}_{a_1\dots a_5;b}  }
{2^{12} \times 5^2}&\nn\\
B^{(3)}_{a_1\dots a_5;b}&=\frac{1}{5}B^{(1)}_{a_1\dots a_5;b}
-\frac{3}{4}B^{(2)}_{a_1\dots a_5;b} &
\end{alignat}
Note that the equation in the last line above must be identically
satisfied, if no new constraints are to arise at dimension zero.
As a consistency check we have verified that this is indeed the
case for the $\nab_a G_{\alpha_1\dots \alpha_4}$ terms in
(\ref{zerodim}).


\section{Conclusion}

In this paper we have examined the theory of deformations of
$D=11$ supergravity from three different perspectives. We have
seen that they can be characterised by elements of certain
spinorial cohomology groups with physical coefficients. The zeroth
order theory can be presented in terms of $\check H^1_F$ or of
$H^{0,3}_F$. In the language of \cite{cntc} these correspond to
the groups ${\cal H}^1, {\cal H}^3$ of tables 3,4 therein. In the
geometrical formulation, the first deformation is given by an
element of $\check H^2_F(phys)$, while in the four-form approach
it is given as an element of $H^{0,4}_F(phys)$. If we include both
the four- and seven-forms we have argued that the first
deformation, assumed to occur at dimension $\ell^6$ and to be at
least quartic in the fields, will be uniquely determined by the
Chern-Simons term needed to cancel the fivebrane anomaly. This
will be true if the cohomology groups $H^{p,q}_{\t}$ all vanish
for $p\geq 3$, which we believe to be the case. Given this, all of
the components of the various superspace field strength tensors
and the modified action including the $R^4$ terms can in principle
be constructed systematically.

It should be noted that the uniqueness of the $R^4$ invariant in
$D=11$ is not purely a result of supersymmetry. We could clearly
construct two Chern-Simons terms corresponding to the two
fourth-order curvature eight-forms which are available and we
could expect to be able to supersymmetrise both of them, which
would imply that both would appear on the right-hand side of the
$G_7$ Bianchi identity. However, the fivebrane anomaly will only
be cancelled by the correct combination of these two terms and so
we must exclude the second superinvariant.

The conclusion that there is just one $R^4$ term in M-theory seems
to be in line with references
\cite{Green:1997di,Green:1997as,Antoniadis:1997eg,Russo:1997mk},
but does not agree with \cite{Tseytlin:2000sf}. According to
\cite{Green:1997di}, in order to obtain the correct
anomaly-cancelling Chern-Simons term, one has to construct a
$D=10, N=2$ superinvariant which includes a particular linear
combination of the two $D=10, N=1$ superinvariants containing CS
terms \cite{deRoo:1992sm,deRoo:1992zp}, and then lift it to
$D=11$.\footnote{The three $N=1, D=10$ $R^4$ invariants discussed
in \cite{deRoo:1992sm,deRoo:1992zp} can easily be understood in
superspace. There are two involving CS terms with the $B$-field to
which can be constructed using the action principle, and the third
is a full superspace integral of any function of the dilaton
superfield, $\f$, which is at least quartic in $\f$.} In
particular, this implies that the appropriately lifted $t_8 t_8
R^4$ term should be a partner of the anomaly-cancelling CS term.
In reference \cite{Tseytlin:2000sf} it is argued that this cannot
be the case because CS terms cannot be ultra-violet divergent in
quantum supergravity, by a background field method argument, while
there is a one-loop divergence in the $t_8 t_8R^4$ term
\cite{Fradkin:1982kf,Green:1997di,Green:1997as,Russo:1997mk}.
However, it is not clear that this argument is
correct.\footnote{This is different to the non-renormalisation of
the Chern-Simons operator into itself in four-dimensional
Yang-Mills theory; in this case the CS operator is a composite
operator which appears multiplied by a source in the
action.\cite{Breitenlohner:1983pi}} Only the action is required to
be gauge-invariant, not the Lagrangian, and this does not seem to
rule out divergences of the CS type. The correct CS term does have
a finite coefficient in M-theory determined by the anomaly, but it
might be the case that this is true even in supergravity. The
five-brane occurs as a soliton in the field theory, and it might
be the case that quantum supergravity knows about this. If this
picture is correct then one could wonder why the one-loop $R^4$
term is infinite, but it could be that the calculated divergence
is in fact due to curvature terms connected to the ``wrong'' CS
multiplet. On the other hand, it could be the case that all of the
coefficients are divergent and that the only thing that
perturbative  quantum supergravity knows about is the
supermultiplet structure. In this case, these coefficients only
get fixed to particular finite values (zero for the wrong
multiplet)  by the anomaly-cancelling mechanism, which could be
said to be necessary both in M-theory and in quantum supergravity
itself.

The question of whether there can be any non-trivial deformations
either with fewer powers of $\ell$ or with fewer than three fields
is one that can be tackled and which we hope to address in the
near future.  In this connection, we note there is another
possible Chern-Simons term which could arise at dimension
$\ell^3$; this would lead to a term of the form $\ell^3 G_4 \tr
R^2$ on the right-hand-side of the $G_7$ Bianchi identity and a
Chern-simons term of the form $C_3 G_4 \tr R^2$. However, such a
term is also incompatible with the five-brane anomaly and should
therefore be discarded.

We note that, although the formalism allows one to determine the
entire theory at order $\b$ in principle, in practice this would
be a very difficult programme to carry through. The
lowest-dimensional non-vanishing component of the twelve-form
$W_{12}$ is $W_{2,10}$ implying that the lowest non-vanishing
component of $K_{11}$ is $K_{3,8}$. Proceeding systematically we
would therefore need to go through eight Bianchi identity steps to
find $K_{11,0}$. If one were to attack the problem this way it
would lead to the complete invariant which would include terms of
up to eight powers of the gravitino field. Clearly this would be a
formidable task. On the other hand, it does not seem to be very
easy to try to solve the problem starting from the highest
dimension and working down. In this case one is nearer to the
purely bosonic part of the invariant which is presumably of most
interest, but there does not seem to be enough information at
first sight to pin it down. This is a problem which needs to be
investigated further.

Finally, one can ask whether the formalism could be used, even in
principle, to go beyond the first deformation. The term we have
found will induce higher-order corrections itself, and these could
in principle be determined. A much simpler example of this is
given in appendix D in the context of deformations of $D=10$ super
Maxwell theory. On the other hand, with more powers of $\ell$
available, one would expect there to be non-vanishing  elements in
$H^{0,7}_F,\, H^{1,6}_F$ and $H^{2,5}_F$ with physical
coefficients and the appropriate dimensions. Although one could
examine these cohomology groups in principle, in practice it is
likely to be prohibitively complicated. One point which might
offer some further constraints on the theory is that there is an
obstruction to any deformation being integrable to the next order.
Again this is briefly discussed for the $D=10$ Maxwell theory in
appendix D. In the context of four-form formulation of
supergravity  the obstruction lies in $H^{0,5}_F(phys)$. It might
be the case that at least some deformations other than those
generated by the fivebrane anomaly are not integrable at higher
order, although this would again be difficult to analyse. We note
that, at sufficiently high dimension ($\del^{12}R^4$) it becomes
possible to write down superinvariants as integrals over the whole
of $D=11$ superspace. It is not clear that supersymmetry alone
places any restrictions on terms of this type.

\section*{Acknowledgements}

This work was supported in part by EU contract HPRN-2000-00122 and
PPARC grants \linebreak PPA/G/S/1998/00613 and PPA/G/O/2000/00451.

We would like to thank K. Stelle for informative remarks,
particularly concerning Chern-Simons terms. We also thank E.
Kiritsis and A. Tseytlin for stimulating discussions.

\section*{Appendix A - the $(n,k)$-tensor projection }

\noindent Consider the (reducible) $(n,k)$-form, $k\leq n$,
$V_{a_1\dots a_n;b_1\dots b_k}$ in $D$ spacetimes dimensions. The
projection onto the irreducible part $\hat{\Pi}^{(n,k)}
V_{a_1\dots a_n;b_1\dots b_k}$ is given by,

\begin{equation}
\hat{\Pi}^{(n,k)} V_{a_1\dots a_n;b_1\dots b_k}=
\sum_{l=0}^{k}(-)^l
\frac{\binom{k}{l}\binom{n+1}{l}}{\binom{D-n-k+l+1}{l}}
\eta_{a_1b_1}\dots \eta_{a_lb_l} F_{(n,k)}^l \label{piproj}
\end{equation}

where,

\begin{equation}
F_{(n,k)}^l:=\frac{n-k+1}{n+1}\sum_{r=0}^{k-l}\binom{k-l}{r}V^{i_1\dots
i_l} {}_{ b_{k-r+1}\dots b_k a_{l+r+1}\dots a_n ;i_1\dots i_l
b_{l+1}\dots b_{k-r}a_{l+1}\dots a_{l+r}}
\end{equation}
Note that in order to simplify the notation, we have omitted the
antisymmetrization brackets. Antisymmetrization is understood in
the $(a_1,\dots a_n)$ and $(b_1,\dots b_k)$ indices. The
normalization is such that $\hat{\Pi}^2=1$. The irreducible part
$\hat{V}:=\hat{\Pi}V$ constructed above satisfies,
\begin{equation}
\hat{V}_{[a_1\dots a_n;b_1]b_2\dots b_k}= \hat{V}^i{}_{a_2\dots
a_n;ib_2\dots b_k}=0.
\end{equation}
In particular we find,
\begin{alignat}{2}
\hat{V}_{a_1a_2;b}
=&\frac{2}{3}(V_{a_1a_2;b}+V_{a_1b;a_2})\nn\\
-&\frac{1}{5}\eta_{ba_1}V^i{}_{a_2;i}
\end{alignat}
\begin{alignat}{2}
\hat{V}_{a_1\dots a_5;b}
=&\frac{5}{6}(V_{a_1\dots a_5;b}+V_{a_1\dots a_4b;a_5})\nn\\
-&\frac{5}{7}\eta_{ba_1}V^i{}_{a_2\dots a_5;i}
\end{alignat}

\begin{alignat}{2}
\hat{V}_{a_1a_2;b_1b_2}
=&\frac{1}{3}(V_{a_1a_2;b_1b_2}+2V_{a_1b_1;a_2b_2}
+V_{b_1b_2;a_1a_2})\nn\\
-&\frac{2}{9}\eta_{b_1a_1}(V^i{}_{a_2;ib_2}
+V^i{}_{b_2;ia_2})\nn\\
+&\frac{1}{45}\eta_{b_1a_1}\eta_{b_2a_2}V^{ij}{}_{;ij}
\end{alignat}
\begin{alignat}{2}
\hat{V}_{a_1\dots a_5;b_1b_2} =&\frac{2}{3}(V_{a_1\dots
a_5;b_1b_2}+2V_{a_1\dots a_4b_1;a_5b_2}
+V_{a_1a_2 a_3b_1b_2;a_4a_5})\nn\\
-&\frac{4}{3}\eta_{b_1a_1}(V^i{}_{a_2\dots a_5;ib_2}
+V^i{}_{a_2a_3 a_4b_2;ia_5})\nn\\
+&\frac{10}{21}\eta_{b_1a_1}\eta_{b_2a_2}V^{ij}{}_{a_3a_4 a_5;ij}
\end{alignat}
\begin{alignat}{2}
\hat{V}_{a_1\dots a_5;b_1\dots b_5} =&\frac{1}{6} (V_{a_1\dots
a_5;b_1\dots b_5}+5V_{a_1\dots a_4b_1;a_5b_2\dots b_5}
+10V_{a_1a_2a_3b_1b_2;a_4a_5b_3b_4b_5}\nn\\
&+10V_{a_1a_2b_1b_2b_3;a_3a_4a_5b_4b_5} +5V_{a_1b_1\dots
b_4;a_2\dots a_5b_5}+V_{b_1\dots b_5;a_1\dots a_5})
\nn\\
-&\frac{5}{3} \eta_{a_1b_1}(V^i{}_{a_2\dots a_5;ib_2\dots b_5}
+4V^i{}_{a_2a_3 a_4b_2;ia_5b_3b_4 b_5}\nn\\
&+6V^i{}_{a_2a_3b_2b_3;ia_4a_5b_4b_5}
+4V^i{}_{a_2b_2b_3b_4;ia_3a_4a_5b_5}
+V^i{}_{b_2\dots b_5;ia_2\dots a_5})\nn\\
+&\frac{25}{6}
\eta_{a_1b_1}\eta_{a_2b_2}(V^{ij}{}_{a_3a_4a_5;ijb_3b_4b_5}\nn\\
&+3V^{ij}{}_{a_3a_4b_3;ija_5b_4b_5}+3V^{ij}{}_{a_3b_3b_4;ija_4a_5b_5}
+V^{ij}{}_{b_3b_4b_5;ija_3a_4a_5})\nn\\
-&\frac{10}{3} \eta_{a_1b_1}\eta_{a_2b_2}\eta_{a_3b_3}
(V^{ijk}{}_{a_4a_5;ijkb_4b_5}\nn\\
&+2V^{ijk}{}_{a_4b_4;ijka_5b_5}+V^{ijk}{}_{b_4b_5;ijka_4a_5})\nn\\
+&\frac{5}{6}
\eta_{a_1b_1}\dots\eta_{a_4b_4}(V^{ijkl}{}_{a_5;ijklb_5}
+V^{ijkl}{}_{b_5;ijkla_5})\nn\\
-&\frac{1}{21} \eta_{a_1b_1}\dots\eta_{a_5b_5}V^{ijklm}{}_{;ijklm}
\end{alignat}

\section*{Appendix B - explicit formulae}

We can use (\ref{piproj}) to give the explicit forms of formulae
(\ref{dia}--\ref{dici}),
\begin{alignat}{3}
DA_{a_1a_2;b_1b_2}=&A^1_{a_1a_2;b_1b_2}&\nn\\
+&\frac{1}{2}(\gamma_{a_1}A^1_{b_1b_2;a_2}+
\gamma_{b_1}A^1_{a_1a_2;b_2})&\nn\\
+&\frac{1}{3}(\gamma_{b_1b_2}A^1_{a_1a_2}+2\gamma_{b_1a_1}A^1_{b_2a_2}
+
 \gamma_{a_1a_2}A^1_{b_1b_2})
\end{alignat}
\begin{alignat}{3}
DB_{a_1\dots a_5;b_1b_2}=&\frac{2}{3}(\gamma_{b_1b_2}B^1_{a_1\dots
a_5}
+2\gamma_{b_1a_1}B^1_{b_2a_2\dots a_5}&\nn\\
&+\gamma_{a_1a_2}B^1_{b_1b_2a_3a_4a_5})
 &\nn\\
+&\frac{2}{5}(\gamma_{b_1b_2a_1}B^1_{a_2\dots a_5}
-2\gamma_{b_1a_1a_2}B^1_{b_2a_3a_4 a_5}&\nn\\
&+\gamma_{a_1a_2a_3}B^1_{b_1b_2a_4 a_5}&\nn\\
&-2\eta_{a_1b_1}\gamma_{b_2}B^1_{a_2\dots a_5}
-2\eta_{a_1b_1}\gamma_{a_2}B^1_{b_2a_3a_4 a_5})&\nn\\
+&\frac{1}{5}(\gamma_{b_1b_2a_1a_2}B^1_{a_3a_4a_5}
+2\gamma_{b_1a_1a_2a_3}B^1_{b_2a_4a_5}&\nn\\
&+\gamma_{a_1a_2a_3a_4}B^1_{b_1b_2a_5}
-4\eta_{b_1a_1}\gamma_{b_2a_2}B^1_{a_3a_4a_5}&\nn\\
&+4\eta_{b_1a_1}\gamma_{a_2a_3}B^1_{b_2a_4a_5}
-\frac{10}{7}\eta_{b_1a_1}\eta_{b_2a_2}B^1_{a_3a_4a_5})\nn\\
+&\frac{1}{15}(\gamma_{b_1b_2a_1a_2a_3}B^1_{a_4a_5}
-2\gamma_{b_1a_1\dots a_4}B^1_{b_2a_5}&\nn\\
&+\gamma_{a_1\dots a_5}B^1_{b_1b_2}
-6\eta_{b_1a_1}\gamma_{b_2a_2a_3}B^1_{a_4a_5}&\nn\\
&-6\eta_{b_1a_1}\gamma_{a_2a_3a_4}B^1_{b_2a_5}
-\frac{30}{7}\eta_{b_1a_1}\eta_{b_2a_2}\gamma_{a_3}B^1_{a_4a_5})\nn\\
+&B^1_{a_1\dots a_5;b_1b_2}&\nn\\
+&\gamma_{a_1}B^1_{a_2\dots a_5b_1;b_2}
+\frac{4}{5}\gamma_{b_1}B^1_{a_1\dots a_5;b_2}&\nn\\
+&\gamma_{a_1a_2a_3}B^1_{a_4a_5;b_1b_2}&\nn\\
-&\frac{1}{5}(\gamma_{a_1\dots a_4}B^1_{b_1b_2;a_5}
+\gamma_{a_1a_2a_3b_1}B^1_{a_4a_5;b_2}&\nn\\
&-\frac{5}{2}\eta_{a_1b_1}\gamma_{a_2a_3}B^1_{a_4a_5;b_2})&\nn\\
+&\dots &
\end{alignat}
\begin{alignat}{3}
DC_{a_1\dots a_5;b_1\dots b_5}= &\frac{1}{6}(\gamma_{a_1\dots
a_5}C^1_{b_1\dots b_5}
+5\gamma_{a_1\dots a_4b_1}C^1_{a_5b_2\dots b_5}&\nn\\
&+10\gamma_{a_1a_2a_3b_1b_2}C^1_{a_4a_5b_3b_4 b_5}
+10\gamma_{a_1a_2b_1b_2b_3}C^1_{a_3a_4a_5b_4 b_5}&\nn\\
&+5\gamma_{a_1b_1\dots b_4}C^1_{a_2\dots a_5 b_5}
+\gamma_{b_1\dots b_5}C^1_{a_1\dots a_5})&\nn\\
+&C^1_{a_1\dots a_5;b_1\dots b_5}&\nn\\
+&\frac{1}{4}(\gamma_{a_1a_2a_3}C^1_{b_1\dots b_5;a_4a_5}
+5\gamma_{a_1a_2b_1}C^1_{b_2\dots b_5a_3;a_4a_5}   &\nn\\
&+5\gamma_{a_1b_1b_2}C^1_{a_2\dots a_5 b_3;b_4b_5}
+\gamma_{b_1b_2b_3}C^1_{a_1\dots a_5;b_4b_5})&\nn\\
+&\frac{1}{5}(\gamma_{a_1\dots a_4}C^1_{b_1\dots b_5;a_5}
+5\gamma_{a_1a_2 a_3b_1}C^1_{b_2\dots b_5a_4;a_5}   &\nn\\
&+10 \gamma_{a_1a_2b_1b_2}C^1_{a_3a_4a_5b_3b_4;b_5}
+5\gamma_{b_1b_2b_3a_1}C^1_{a_2\dots a_5b_4;b_5}
&\nn\\
&+\gamma_{b_1\dots b_4}C^1_{a_1\dots a_5;b_5} )&\nn\\
+&\dots &
\end{alignat}

\section*{Appendix C: Representation-theoretical conventions }

Consider a Lie group $G$ and a $G$-module $V$. The $n$-th tensor
product $V^{\otimes n}$ admits a decomposition
$$
\sum_{R} V_R \times R
$$
under $G\times S_n$, where $R$ runs over all irreducible
representations of the symmetric group $S_n$ and $V_R$ is a
$G$-module. As is well known, the irreps of $S_n$ can be
parametrized by partitions of $n$ or, equivalently, by the
associated Young diagrams. If $R$ is associated to the partition
$\l$ of $n$, $V_R$ is the {\it plethysm} of $V$ with respect to
$\l$.

For the case at hand, the highest weight of the standard module
(i.e. of the vector representation) of $Spin(1,10)$ is given by
$(10000)$ on the basis of fundamental weights. A $k$-form is
associated to the partition $[1,\dots ,1]$ of $k$. Its highest
weight for $k\leq 4$ is $(0\dots 010\dots 0)$, with the nonzero
entry on the $k$-th position. The five-form is represented by
$(00002)$, as the $(00001)$ is the highest weight of the spinor
representation. Moreover, the highest weight of the
$k$-form-spinor is given by the sum of the highest weight of the
$k$-form and the highest weight of the spinor. Recall that by a
$k$-form-spinor we refer to the projection onto the irreducible
(gamma-traceless) part. For example $(10001)$ is a vector-spinor,
$(01001)$ is a two-form-spinor, etc. We can also consider the
projection onto the highest-weight representation of the plethysm
associated to the partition $[2,\dots  ,2,1,\dots ,1]$ of $n+k$,
with $k$ entries equal to 2 and $n-k$ entries equal to 1. This
provides an alternative way to view the $(n,k)$-tensor discussed
in Appendix A. Note that $n,k\leq 5$ in the case of $Spin(1,10)$.
The highest weight of an $(n,k)$-tensor with $k,n< 5$ is given by
$(0\dots 010\dots 010\dots 0)$ where the nonzero entries are on
the $k$-th and $n$-th positions. For $k<n=5$ the highest weight is
$(0\dots 010\dots 02)$ with the first nonzero entry on the k-th
position. Finally, for $k=n=5$ the highest weight is $(00004)$. In
other words, the highest weight of the $(n,k)$-tensor is given by
the sum of the highest weight of a $k$-form and the highest weight
of an $n$-form. The highest-weights of irreducible
(gamma-traceless) $(n,k)$-tensor-spinors are represented in the
obvious way. For example $(01101)$ is the highest weight of a
$(3,2)$-tensor-spinor, $(10003)$ that of a $(5,1)$-tensor-spinor,
etc. This discussion generalizes straightforwardly to
representations corresponding to more general partitions.



\section*{Appendix D: cohomology}

\subsection*{Pure spinors}

Pure spinors were introduced in $D=10$ super Maxwell theory in an
attempt to find an off-shell version \cite{Nilsson:1985cm}.
Subsequently it was pointed out that the on-shell constraints of
theories in ten and eleven dimensions can be interpreted in terms
of pure spinor integrability \cite{Howe:mf,Howe:1991bx}}. A pure
spinor in $D=10$ is a complex bosonic chiral spinor $u^\a$ obeying
the single constraint $u^\a (\c^a)_{\a\b} u^\b=0$. Pure spinor
integrability implies that $u^\a u^\b F_{\a\b}=0$ where $F=dA$ is
the super Maxwell field strength. In the notation of the current
paper this can be written as $[F_{0,2}]=0$ and this in turn
implies that $d_F[A_{0,1}]=0$. Taking gauge transformations this
means that the field equations of $D=10$ super Maxwell are encoded
in the cohomology group $H^{0,1}_F$. Berkovits noticed this in his
work and rephrased it slightly differently
\cite{Berkovits:2001rb}. He assigns ghost number 1 to the pure
spinor $u$ and introduces the BRST operator $Q:=u^\a D_{\a}$ which
squares to zero by virtue of the pure spinor constraint. The
cohomology group $H_Q^p$ is defined to be functions $A$ which have
ghost number $p$ and which are $Q$-closed modulo $Q$-exact ones. A
typical function looks like

 \be
 A=A_o + u^\a A_\a + u^\a u^\b A_{\a\b} + \ldots
 \ee

where the multispinors in the expansion are all symmetric and
gamma-traceless. It is immediately clear that that $H_Q^p$ is
isomorphic to $H^{0,p}_F$. In the Maxwell case Berkovits showed
that the only non-trivial groups have $p=0,1,2,3$ with $H_Q^0\sim
H_Q^3$ and $H_Q^1\sim H_Q^2$. The elements of $H_Q^1$ defined the
physical fields of the theory as we have seen, and the elements of
$H^2_Q$ can be interpreted as the corresponding anti-fields. We
also find that $H^0_Q=H^3_Q=\bbR$ and these two groups are related
to the gauge ghost and its anti-field.

\subsection*{Spinorial cohomology}

The notion of spinorial cohomology was introduced independently by
the G\"oteborg group and used in studies of super Yang-Mills and
supergravity theories \cite{cntc,cntb,cntd}. The concept turns out
to be isomorphic to pure spinor cohomology but is presented in a
slightly different fashion which is helpful in computing the
representations that arise. We illustrate the idea in the $D=10$
Maxwell case. One considers a complex of representations $\{r_k\}$
of the spin group consisting of totally symmetric, gamma-traceless
$k$-spinors. (Clearly the same as Berkovits's pure spinor
wave-functions). However, one then tensors each such
representation with the $l^{th}$ antisymmetric  power of the
fundamental spinor to arrive at an array of representations
$\{r^l_k\}$. The representations contained in the element $r^l_k$
are the representations that occur in the $\th^l$ component of a
superfield whose leading component is in the representation $r_k$.
The original set of representations is made into a complex by
differentiating a superfield with the spinorial derivative and
then projecting onto the gamma-traceless part of the result.
Clearly the cohomology obtained this way is isomorphic to pure
spinor cohomology if one uses the supercovariant derivative, but
if one simply wants to know which representations are present one
can use the partial spinorial derivative. It is easy to see that
this induces a map $r^l_k\rightarrow r^{l-1}_{k+1}$, and this fact
can be used to compute the cohomology using group-theoretic
techniques. If we let an element of $r^l_k$ be denoted by
$J_{\a_1\ldots \a_k,\b_1 \ldots \b_l}$, then the induced operation
is

 \be
 J_{\a_1\ldots \a_k,\b_1 \ldots \b_l}\mapsto
 J_{\{\a_1\ldots \a_k,\b_1\} \ldots \b_l}\in r^{l-1}_n
 \ee

where the braces denote gamma-traceless symmetrisation. It is
apparent that this defines a linear map $\d:r^l_k\rightarrow
r^{l-1}_{k+1}$ and that $\d^2=0$. This resembles the Spencer
cohomology \cite{spencer} of a bi-complex $\{C_{n,l}\}$ of tensors
which are symmetric on the first $n$ indices and antisymmetric on
the second $l$ with operators $\d:C_{k,l}\rightarrow C_{k+1,l-1}$
and $\d': C_{k,l}\rightarrow C_{k-1,l+1}$ defined in the obvious
way. In this case since $\d\d'+\d'\d=1$ the cohomology is trivial,
but in our case this conclusion is avoided by the projection onto
the gamma-traceless multi-spinors.

\subsection*{$D=10$ super Maxwell}

To illustrate the meaning of the cohomology groups we again look
at $D=10$ SYM. The two lowest-dimensional components of the
Bianchi identity are

 \bea
 d_1 F_{0,2} + \t_0 F_{1,1}&=&0 \\
 d_1 F_{1,1} + \t_0 F_{2,0}&=&0
 \eea

As we have already noted, the on-shell (free) theory is obtained
by setting $[F_{0,2}]=0$ and this is equivalent to specifying an
element of $H^{0,1}_F$. In this case we note that we can set
$F_{0,2}=0$ in which case $F_{1,1}$ determines an element of
$H^{1,1}_F$. This gives an alternative cohomological way of
describing the theory. One can easily see that this cohomology
group is a spinor superfield $\l^\a$ satisfying the constraint
$D_\a \l^\b = (\c^{ab})_\a{}^\b F_{ab}$, i.e. it is the on-shell
field strength supermultiplet.

Now consider the group $H^{0,2}_F$. We can consider the equation
of motion of the theory to be $F_{\{\a\b\}}=0$; if we relax this
by introducing a current $J_{\{\a\b\}}$ on the right-hand side
then the latter must be spinorially closed by virtue of the
Bianchi identity. On the other hand, if it is trivial, the
connection can be redefined to regain the original equations of
motion. This implies that $H^{0,2}_F$ describes the currents of
the theory which are in one-to-one correspondence with the
anti-fields. But now suppose that we are interested in looking at
deformations of the theory. In this case we need to look at the
same cohomology group but the coefficients will be restricted to
be tensorial functions of the field strength superfield $\l$ and
its derivatives. We denote this group by $H^{0,2}_F(phys)$. The
two groups are quite different; the former is dual to the physical
fields while the latter describes a composite multiplet in the
theory.

If we deform the theory with respect to a parameter $t$ (in
practice $\a'^2$) then the above discussion is relevant to the
first deformation. In fact we need only have
$d_F[\Fo_{0,2}]=t[\chi_{0,3}]+ O(t^2)$, where the function $\chi$
is calculated using the first-order corrections to the equations
of motion. This term has to be taken into account when we go to
second order; it is clearly closed but it also needs to be exact
in order for it to be absorbed into the second order deformation.
This was observed in \cite{cntd}.  From this discussion it is
clear that the possible obstruction to this lies in
$H^{0,3}_F(phys)$. Provided that the class so defined is trivial,
this term, and similar terms at higher order, will give rise to
correction terms in the equations of motion, or the component
Lagrangian, which are induced by the first deformation. For super
Maxwell theory we know that Born-Infeld provides a consistent
deformation to all orders, so it would be interesting to see how
much of the Born-Infeld theory is induced by the first deformation
as a result of $N=1$ supersymmetry in ten dimensions. Indeed, it
is known that the $F^6$ term is induced from the $F^4$ term in
this way \cite{Collinucci:2002gd}.

\subsection*{Spinorial cohomology another way}

Here we give a brief account of spinorial cohomology in superspace
in a rather general setting making as few assumptions as possible.
Let $M$ denote superspace. We shall assume only that it is
equipped with a choice of odd tangent bundle $F$ and that the
Frobenius tensor $\vf$ defined below gives a projection from
$\wedge^2 F$ to the even tangent bundle $B$ which at this stage we
can take to be the qotient bundle $T/F$. For any two sections
$X,Y\in \C( F)$ the Frobenius tensor is defined by

\be \vf(X,Y) =[X,Y]\ {\rm mod} \ F \ee

If we introduce a local basis of vector fields $\{E_{\a}\}$ for
$F$ and a local basis of one-forms $\{E^a\}$ for the even
cotangent space $B^*$, the components of $\vf$ with respect to
this basis are essentially the same as those of the dimension-zero
component of the torsion tensor.

We let $\O^p$ denote the space of $p$-forms, and $I$ the ideal of
forms generated by $B^*$, i.e.the space of all forms which have at
least one even index. Set $\O^p_F=\O^p/I^p$ where $I^p:=I\cap
\O^p$. This space can be thought of as the space of purely odd
(fermionic) $p$-forms. We would like to define an exterior
derivative which maps $\O^p_F$ to $\O^{p+1}_F$ but this is not
possible due to the fact that two odd derivatives do not
anticommute because of the requirement of supersymmetry. We
therefore define the space of reduced odd $p$-forms, $\hat\O^p_F$,
by

\be \hat\O^p_F:=\O^p/ \left(I^p \cup d I^{p-1}\right) \ee

The point of this definition is to get rid of the unwanted terms
arising from the anticommutator. It is straightforward to define a
derivative $d_F$ as follows: let $[\o_p]\in \hat\O^p_F$ where
$\o_p\in \O^p$ is a representative of this equivalence class, then
set

\be d_F [\o_p]:=[d\o_p]  \ee

It is easy to check that this definition is independent of the
choice of representative. If $\o_p\mapsto \o_p + \l_p +
d\m_{p-1}$,  with $\l_p\in I^p$ and $\m_{p-1}\in I^{p-1}$, then
$d\o_p\mapsto d\o_p + d \l_p$. But the second term here is an
element of $dI^p$ which is factored out in $\hat\O^{p+1}_F$. It is
also trivial to see that $d_F^2=0$.

This construction can be extended to $(p,q)$ forms and to
$B$-valued odd forms.


\end{document}